\documentclass[12pt, referee]{aastex}
\usepackage[dvips]{color}

\newcommand{\bu}{\textcolor[rgb]{0,0,0}}
\newcommand{\ma}{\textcolor[rgb]{0.0,0,0.0}}
\newcommand{\gr}{\textcolor[rgb]{0,0.0,0}}
\newcommand{\be}{\textcolor[rgb]{0,0,0}}
\newcommand{\bx}{\textcolor[rgb]{0,0,0}}
\newcommand{\bz}{\textcolor[rgb]{0,0,0}}
\newcommand{\bzz}{\textcolor[rgb]{0,0,0}}
\newcommand{\bzzz}{\textcolor[rgb]{0,0,0}}
\newcommand{\byy}{\textcolor[rgb]{0,0,0}}
\newcommand{\bzzzz}{\textcolor[rgb]{0,0,0}}
\def\gtsima{$\;\buildrel > \over \sim \;$}
\def\simgt{\lower.5ex \hbox{\gtsima}}
\def\ltsima{$\;\buildrel < \over \sim \;$}
\def\simlt{\lower.5ex \hbox{\ltsima}}

\begin{document}

\title{Dark Matter that Interacts with Baryons: Density Distribution within the Earth and New Constraints on the Interaction Cross-section}
\author{David A. Neufeld\altaffilmark{1}, Glennys R.\ Farrar\altaffilmark{2}, and Christopher F. McKee\altaffilmark{3}}

\altaffiltext{1}{Department of Physics \& Astronomy, Johns Hopkins University, Baltimore, MD 20218, USA}
\altaffiltext{2}{Center for Cosmology and Particle Physics, Department of Physics, New York University, New York, NY 10003, USA}
\altaffiltext{3}{Department of Physics and Department of Astronomy, University of California, Berkeley, CA 94720, USA}

\begin{abstract}

For dark matter (DM) particles with masses in the $0.6 - 6 m_{\rm p}$ range, we set stringent constraints on the interaction cross-sections for scattering with ordinary baryonic matter. These constraints follow from the recognition that such particles can be captured by - and thermalized within - the Earth, leading to a substantial accumulation and concentration of DM that interact with baryons. Here, we discuss the probability that DM intercepted by the Earth will be captured, the number of DM particles thereby accumulated over Earth's lifetime, the fraction of such particles retained in the face of evaporation, and the density distribution of such particles within the Earth. \byy{In the latter context, we note that a previous treatment of the density distribution of DM, presented by Gould and Raffelt and applied subsequently to DM in the Sun, is inconsistent with considerations of hydrostatic equilibrium.} Our analysis provides an estimate of the DM particle density at Earth's surface, which may exceed $10^{14}\,\rm cm^{-3}$, and leads to constraints on various scattering cross-sections, which are placed by: (1) the lifetime of the relativistic proton beam at the Large Hadron Collider; (2) the orbital decay of spacecraft in low Earth orbit; (3) the vaporization rate of cryogenic liquids in well-insulated storage dewars; and (4) the thermal conductivity of Earth's crust. For the scattering cross-sections that were invoked recently in Barkana's original explanation for the anomalously deep 21 cm absorption reported by EDGES, DM particle masses in the $0.6 - 4 m_{\rm p}$ range are excluded.

\end{abstract}


\section{Introduction}

In recent years, there has been considerable interest in the possibility that dark matter (DM) might have astrophysically-significant non-gravitational interactions with ordinary baryonic matter, 
\be{prompted in part by the suggestion that DM may be self-interacting (e.g.\ Spergel \& Steinhardt 2000)}.  The possible effects of such interactions have been discussed in a wide variety of astrophysical contexts, including \be{the heating of X-ray
clusters (Qin \& Wu 2001; Chuzhoy \& Nusser 2006)}, big bang nucleosynthesis (BBN; e.g.\ \gr{Cybert et al.\ 2002}); the power spectrum of the cosmic microwave background (CMB) \be{ and Lyman-$\alpha$ forest} (e.g.\ \gr{\be{Dvorkin} et al.\ 2014; \bz{Xu et al.\ 2018}; Gluscevic \& Boddy 2017}); and the thermal balance of neutral atomic gas during the ``Dark Ages" (Mu{\~n}oz et al.\ 2015) and at ``Cosmic Dawn" (Barkana 2018; Fialkov et al.\ 2018; Mu{\~n}oz) \& Loeb 2018). \be{Additional effects result if DM can annihilate with ordinary matter (e.g.\
Farrar \& Zaharijas 2006; Mack et al.\ 2007)}. Such considerations set constraints on the allowable cross-sections for scattering with ordinary baryonic matter, as a function of the particle mass.  \be{For cross sections large enough to have astrophysical effects, direct detection
searches for WIMP dark matter (e.g.\ Abrams et al.\ 2002; Aguilar-Arevalo et al.\ 2016;
Angloher et al. 2017) do not generally apply, because DM particles lose too much energy
via collisions in the atmosphere or Earth's crust to trigger the detector (e.g.\ Starkman
et al.\ 1990). Numerous studies have been performed to place limits for the ``moderately
interacting" range of cross sections, starting with Wandelt et al.\ 2001;
\bx{an extensive discussion of the limits from direct detection 
experiments has been presented very recently by Mahdawi \& Farrar (2018).}}
Each of the various constraints thereby obtained typically applies to specific baryonic nuclei and for a specific range of collision velocities.  A wide variety of interactions have been considered, ranging from interactions involving hadronic dark matter (e.g.\ a stable sexaquark; Farrar 2017) that are characterized by a Yukawa (or double-Yukawa potential); to long-range interactions --  involving as-yet undiscovered forces \be{or milli-weak
charged DM} -- with a cross-section that is posited to decrease rapidly with collision velocity (e.g. Mu{\~n}oz et al.\ 2015), as in the case of Rutherford scattering.

In this paper, we consider additional constraints that can be derived in the circumstance that DM particles are captured by -- and concentrated within -- the Earth.   The concentration of DM within the Earth can be significant within the range of parameters to be considered in the present paper:
DM masses, $m_{\rm DM}$, between $0.2$ and $10\,m_{\rm p}$, and scattering cross-sections in the range $10^{-30}$ to $10^{-20}\,\rm cm^2$ for typical nuclei in the crust or atmosphere.  
To avoid confusion with ``self-interacting dark matter" (SIDM, introduced by Spergel \& Steinhardt), we adopt the acronym HIDM, for ``hadronically-interacting dark matter," to refer to particles in this region of parameter space.

In Section 2, we discuss the capture of HIDM particles by the Earth, which follows the scattering and thermalization of HIDM in Earth's crust or atmosphere.  In Section 2, we discuss the probability that a HIDM particle that is intercepted by the Earth will be captured\be{,} the number of HIDM particles thereby accumulated over Earth's lifetime\be{,} the fraction of such particles retained the face of evaporation (``Jeans loss")\be{,} and the density distribution of such particles within the Earth.  These considerations lead to an estimate of the DM particle density at Earth's surface, which we use in Section 3 to determine constraints on the scattering cross-sections.  These constraints are placed by four considerations: (1) the lifetime of the relativistic proton beam at the Large Hadron collider (LHC); (2) the orbital decay of spacecraft in low Earth orbit (LEO);
(3) the vaporization rate of cryogenic liquids in well-insulated storage dewars; and (4) the thermal conductivity of Earth's crust.  In Section 4, we present a summary of these combined constraints.

\section{Density distribution of hadronically-interacting DM within the Earth}

\subsection{Scattering of HIDM within Earth's atmosphere and crust}

As it moves through the Galaxy, the Earth intercepts DM matter along its path.   An incoming HIDM particle striking the Earth at normal incidence suffers its first scattering at a typical distance $z_s$ above Earth's surface, where
$$\int_{z_s}^\infty {dz \over \lambda} = \int_{z_s}^\infty d\tau = 1, \eqno(1)$$
Here, $\lambda$ is the mean free path for scattering, which depends upon the 
average scattering cross-section for those nuclei that are present, \bx{and $\tau$ is the optical depth, i.e.\ the number of interaction lengths traversed.}
\ma{We denote the cross-section for the scattering of HIDM with an atomic nucleus A} \gr{at relative velocity $v$ as $\sigma^{\rm A}_{\rm v}$.  \bzzz{Here, and henceforth in this paper, the cross-sections we refer to are always the momentum-transfer cross-sections, $\int (d\sigma/d\Omega) (1-\cos\theta) d\Omega$, where $\theta$ is the scattering angle in the center-of-mass frame and $d\sigma/d\Omega$ is the differential scattering cross-section.  In general, the latter is function of $\theta$.  If $d\sigma/d\Omega$ is an even function of $\cos \theta$ (i.e. if the scattering has a forward-backwards symmetry), then the momentum-scattering cross-section is equal to the total scattering cross-section, $\int (d\sigma/d\Omega) d\Omega$.}  
The total vertical column density of atoms in Earth's atmosphere is $P_{\rm sl}/(g{\bar m}_{\rm A}) \sim 4 \times 10^{25}\, \rm \, cm^{\bx{-2}}$, where $P_{\rm sl} \sim 1.0 \times 10^6\,\rm dyne \, cm^{\bx{-2}}$ is the pressure at sea-level, $g \sim 980\, \rm cm\, s^{-1}$ is the acceleration due to gravity, and ${\bar m}_{\rm A} \sim 14.5\,\bu{m_{\rm p}}$ is the mean atomic mass.}  
Thus, if the average cross-section is greater than $\sim \gr{2.5} \times 10^{-26}\,\rm cm^{2}$, incoming HIDM will scatter first in Earth's atmosphere ($z_s\ge 0$); here, the elemental composition (at sea-level) is $\sim 23.2\%$ O, $75.5\%$ N, and $1.3\%$ Ar by mass.  But for average cross-sections within the $3 \times 10^{-29}$ to $2.5 \times 10^{-26}\,\rm cm^2$ range, HIDM first scatter within the top 5~km of the crust ($\sim -5 {\, \rm km} \le z_s \le  0$); for the crust, the typical elemental composition by mass is $46.6\%$ O, $27.7\%$ Si, $8.1\%$Al, 5.0$\%$ Fe, 3.6$\%$ Ca, 2.8$\%$ K, 2.6$\%$ Na, 1.5$\%$ Mg, and 1.5$\%$ other elements.  

\ma{The variation of the scattering cross-section, $\sigma^{\rm A}_{\rm v},$ from one nucleus to another depends upon the nature of the scattering potential.  For a interaction involving a long-range $(1/r$) potential proportional to nucleon number, the dependence is 
$\sigma^{\rm A}_{\rm v} \propto v^{-4} (A/\mu_{\rm A})^2$, where
$$\mu_{\rm A}= Am_{\rm p} m_{\rm DM}/(Am_{\rm p} + m_{\rm DM}) \eqno(2)$$ 
is the reduced mass.  This behavior is exactly analogous to that obtained for scattering in a Coulomb potential (i.e.\ Rutherford scattering):  
$\sigma^{\rm A}_{\rm v} \propto v^{-4} (Z/\mu_{\rm A})^2,$ where $Z$ is the nuclear charge. But for a short-range interaction, such as that described by a Yukawa potential, the situation is more complicated.  In the Born approximation, the cross-section for this case}
is given by a simple dependence that has been widely used in previous studies:  $\sigma^{\rm A}_{\rm v} \propto A^2 \mu_{\rm A}^2$ (e.g.\ Kurylov \& Kamionkowski 2004).  While this simple $A$-dependence might apply at high collision energies \bzzz{or if the coupling is weak}, very substantial deviations could occur at low energies and might result in a scattering cross-section that shows a strong and non-monotonic dependence on $A$. Such is indeed the case for the scattering of thermal neutrons by atomic nuclei.  In particular, an attractive Yukawa potential can give rise to resonances associated with quasi-bound states; this behavior opens the possibility that $\sigma^{\rm A}_{\rm v}$ might be greatly enhanced or reduced for specific values of $A$ (Farrar \& Xu 2018).   In this study, we \be{allow for} the possibility that $\sigma^{\rm A}_{\rm v}$ varies rapidly with $A$ in a manner poorly-described by the Born approximation.  Accordingly, we will present limits on $\sigma^{\rm A}_{\rm v}$ for a variety of different nuclei and for a variety of mixtures of nuclei.  \ma{In the notation we adopt here, the superscript on $\sigma^{\rm A}_{\rm v}$ is either the chemical symbol for the nucleus in question, or denotes a medium -- e.g. the crust, denoted by a superscript ``cr" -- for which we constrain the average scattering cross-section for the various nuclei it contains.  Where a number appears alone in the subscript, it is the collision velocity in $\rm km\,s^{-1}$.  Alternatively, the subscript may denote a collisional energy (e.g.\ ``6.5~TeV'') or a temperature (e.g.\ ``300  K") for which the velocity-averaged cross-section is constrained.}

\subsection{Number of HIDM particles captured by Earth}

We consider first the capture of halo HIDM by the Earth.  During its lifetime, the number of HIDM particles intercepted by the Earth is 
$$N_{\rm I}= {t_\earth \rho_{\rm DM} v_\earth \pi R_\earth^2 \over m_{\rm DM}} = 1.16 \times 10^{42} \biggl({\rho_{\rm DM} \over 0.3\,(\rm{GeV}/c^2) \,{\rm cm}^{-3}}\,\biggr) \biggl({v_\earth \over {\rm 200\,km\,s^{-1}}} \biggr) \biggl({m_{\rm DM} \over {m_{\rm p}}}\biggr)^{-1}, \eqno(3)$$
where \ma{$t_\earth$ = 4.55~Gyr (Manhes et al.\ 1980) is the age of the Earth}, $\rho_{\rm DM}$ is the dark matter mass density in the Galactic plane, $v_\earth$ is the average velocity of the Earth relative to the HIDM particles, and $R_\earth = 6371$~km is the radius of the Earth.  On the right-hand side of equation (3), we have normalized $v_\earth$, $m_{\rm DM}$, and $\rho_{\rm DM}$, relative to typical values of interest for the rotational velocity of the Sun around the Galactic Center, the HIDM mass, and the DM density in the Galactic plane.  
\ma{In an extensive review of the local density of dark matter, Read (2014) found that 
determinations of $\rho_{\rm DM}$ the two years prior to the review were in the range 
$0.2-0.8\, (\rm{GeV}/c^2) \,{\rm cm}^{-3}$  (excluding one non-detection). Subsequent analyses have led to values $\rho_{\rm DM}\simeq 0.5\, (\rm{GeV}/c^2) \,{\rm cm}^{-3}$ (Bienayme et al.\ 2014; Piffl et al.\ 2014;  McKee et al.\ 2015, and Sivertsson et al.\ 2017). Nonetheless, in order to be conservative, we adopt a lower value, $\rho_{\rm DM}=0.3 \,(\rm{GeV}/c^2) \,{\rm cm}^{-3}$, for purposes of our analysis.  \bzzzz{All the upper limits obtained in Section 3 below on the cross-sections for the interaction of HIDM with ordinary matter are inversely proportional to the value adopted for $\rho_{\rm DM}$, while the lower limits obtained in Section 3.4 are proportional to $\rho_{\rm DM}$.  Thus, larger values of $\rho_{\rm DM}$ than the conservative value we adopt would strengthen all the constraints obtained in  our study.}
Equation (3) is based upon the assumption that the Earth presents a cross-section $\pi R_\earth^2$ to halo DM; this neglects small enhancements that might result from the thickness of the atmosphere (in the case where DM first scatter in the atmosphere), and from gravitational focusing effects (which are small because in typical models for the DM halo $v_\earth$ is much larger than the Earth's escape velocity.)} \bu{Equation (3) also neglects the effects of random DM motions that \be{would lead} to modest enhancements in the rate at which DM hit the Earth, and \be{ignores possible evolution in the local DM density and velocity distribution over the lifetime of the Earth.}}
 
Not every intercepted particle is captured, however, because a significant fraction of the particles is reflected from the Earth's atmosphere with a speed larger than Earth's escape velocity, $v_{\rm es} = 11.2\rm\,km\, s^{-1}$.  The fate of any incoming HIDM particle depends upon the number of scatterings it suffers before emerging again from the atmosphere: a particle that suffers a sufficient number of scatterings will lose enough energy to emerge with a speed smaller than $v_{\rm es}$ and will thus be captured.

When a fast DM particle scatters off a much slower atom of mass $m_{\rm A}$ in Earth's atmosphere, the fraction of kinetic energy transferred is $$f_{\rm KE} = {2 (1 - \cos \theta) m_{\rm DM} m_{\rm A} \over (m_{\rm DM} + m_{\rm A})^2}, \eqno(4)$$ where $\theta$ is the scattering angle in the center-of-mass frame.  In this expression for $f_{\rm KE}$, the relevant mass is that of the scattering nucleus, not the atmospheric molecule that contains it, because energy transfer to the scattering nucleus occurs on a short timescale relative to the rotational and vibrational periods of the molecule. If the scattering cross-section has a forwards-backwards symmetry, then the mean kinetic energy transfer is
${\bar f_{\rm KE}} = 2m_{\rm DM} m_{\rm A} / (m_{\rm DM} + m_{\rm A})^2,$  
and the mean number of scatterings needed to reduce the particle speed below the escape velocity is $N_0 = \ln (v_{\rm es}^2 /v_\earth^2) / \ln (1-{\bar f_{\rm KE}})$.  For example, a particle of mass $m_{\rm DM} = 2m_{\rm p}$ and velocity $\rm 200\,km\,s^{-1}$ scattering off pure nitrogen atmosphere ($m_{\rm A} = 14m_{\rm p}$) has $\bar f_{\rm KE} = 7/32$ and $ N_0= 23.3$.  

The problem to be addressed here -- determining the fraction of particles that suffer a given number of scatterings before reflection from a scattering slab -- is almost exactly analogous to an astrophysical problem that \be{was} discussed more than three decades ago: the reflection of X-rays by a cloud of cold electrons (Lightman \& Rybicki 1979; Lightman et al.\ 1981, hereafter L81).  In this case, when the number of scatterings exceeds $\sim 5$, the fraction of reflected X-ray photons to have suffered $N$ scatterings is well-approximated by $\pi^{-1/2} N^{-3/2}$ (L81, their equation 14)\footnote{A simple argument can explain the $N^{-3/2}$ dependence obtained by L81.  If $\tau$ represents the penetration depth of incident particles in units of the mean free path, then the average depth reached by an incident particle before the first scattering is $\tau \sim 1$.  After scattering at $\tau = 1$, roughly one-half of the particles are reflected (i.e. reach $\tau=0$) prior to reaching depth $\tau = 2$, and roughly one-half reach depth $\tau = 2$.  Of those particles that reach depth $\tau = 2$, we may argue (by symmetry) that roughly one-half are reflected (i.e. reach $\tau=0$) without ever reaching depth $\tau = 4$, and roughly one-half reach depth $\tau = 4$ or deeper.  This argument then implies that the fraction of particles that penetrate to depth $\tau$ or deeper is of order $\tau^{-1}$.  But, in a random walk process, the number of scatterings suffered by particles that penetrate to depth $\tau$ or deeper is of order $\tau^2$ or larger.  Thus the fraction of particles that suffer $N$ or more scatterings is of order $N^{-1/2}$, and the fraction that suffer exactly $N$ scatterings is of order $N^{-3/2}$.}.  Thus, by analogy, the fraction of incident HIDM particles that suffer enough scatterings (i.e.\ $N_0$ or more) to have their speed reduced below $v_{\rm es}$ and be captured is
$$f_{\rm cap} = 2 \pi^{-1/2} N_0^{-1/2} = 2 \pi^{-1/2} [\ln (v_{\rm es}^2 /v_\earth^2) / \ln (1-{\bar f_{\rm KE}})]^{-1/2}.\eqno(5)$$  
For the example considered previously (i.e.\ 2$m_{\rm p}$ particles at velocity $\rm 200\,km\,s^{-1}$ scattering in the atmosphere), $f_{\rm cap} = 0.23.$  

\subsection{Density of captured particles within the Earth}

We defer to Section 2.4 a discussion of the loss of HIDM from the atmosphere or the crust.  Under conditions where the loss rate is negligible, the average number density within the Earth is
$${\bar n_{\rm DM}} = {3 f_{\rm cap} N_{\rm I} \over 4 \pi R_\earth^3} = {3 f_{\rm cap} t_\earth \rho_{\rm DM} v_\earth \over 4 R_\earth m_{\rm DM}} $$ 
$$= 2.46 \times 10^{14} \biggl({f_{\rm cap} \over 0.23} \biggr) \biggl({\rho_{\rm DM} \over 0.3\,(\rm{GeV}/c^2) \,{\rm cm}^{-3}}\,\biggr) \biggl({v_\earth \over {\rm 200\,km\,s^{-1}}} \biggr) \biggl({m_{\rm DM} \over {m_{\rm p}}}\biggr)^{-1} \rm \, cm^{-3}. \eqno (6)$$

Provided that the mean free path within the Earth, $\lambda$, is much smaller the length scale on which the temperature varies, the HIDM particles will reach thermal equilibrium \ma{with local material in the Earth} and will acquire a Maxwell-Boltzmann velocity distribution with a mean speed $\bar v = (8kT/[\pi m_{\rm DM}])^{1/2} = 2.51 (T/300\,{\rm K})^{1/2} (m_{\rm DM}/m_{\rm p})^{-1/2}\,\rm km\,s^{-1}.$  The HIDM particles captured by the Earth will assume an equilibrium density distribution on a characteristic diffusion timescale, $t_{\rm diff} \sim R_\earth^2 / (\lambda {\bar v})$, that is typically short compared to the age of the Earth.  For the ${\bar v}$ \be{in} the Earth's core, where the product of $\lambda {\bar v}$ is smallest, we may obtain an upper limit $t_{\rm diff} \le 1.1 \times 10^{4} (m_{\rm DM}/m_{\rm p})^{-1/2} (\lambda/{\rm cm})^{-1}\, \rm yr$.   

Provided $t_{\rm diff} \ll t_\earth$, the number density of HIDM particles within the Earth, $n_{\rm DM}$, is governed by the Jeans equation (e.g.\ Binney \& Tremaine \be{2008}), 
which -- for a static, steady-state distribution of particles  -- simplifies to
$$ \sum_i {\partial (n_{\rm DM} \sigma_{ij}^2) \over \partial x_i} = -n_{\rm DM} {\partial \Phi \over \partial x_j}, \eqno(7)$$ where $\Phi$ is the gravitational potential and $\sigma_{ij}^2 = <v_i v_j>$ is the velocity dispersion tensor.  Here, we may neglect the effects of Earths' rotation about its axis and its motion around the Sun, because the accelerations associated with those motions are much smaller than the acceleration due to Earth's gravity. \bz{Equation (7), which applies to both collisionless and collisional gases, may be derived from the collisional Boltzmann equation, as shown by Chapman \& Cowling (1970, hereafter CC70; their \bzz{Chapter 8}), who present a derivation of the divergence of the pressure tensor (their equation \bzz{8.1,7) for a single component within a gas mixture.}  For $t_{\rm diff} \ll t_\earth$, the HIDM particles have reached their final density distribution and have no net motion relative to the Earth, i.e.\ their average velocity, $\bar{C_1}$ (in the notation of CC70) is zero.  In this case, CC70 equation (8.1,7) reduces to equation (7) above, after division by the particle mass.}

\bzzz{If the length scale on which the temperature changes, $-(d\ln T/dr)^{-1}$, is large compared to the thermalization length, $\lambda_*$, then the velocity distribution is isotropic and in thermal equilibrium with local material in the Earth: $\sigma_{ij}= (kT/m_{\rm DM})^{\bz{ 1/2}} \delta_{ij}.$  Here, the thermalization length (e.g. Rybicki \& Lightman 1986, p.~38) is the root mean square radial distance traveled by an HIDM particle before reaching 
thermal equilibrium with its surroundings.  If the mean fractional energy transfer is ${\bar f}_{\rm KE}$ per scattering, then of order ${\bar f}_{\rm KE}^{-1}$ scatterings are required for thermalization.  During thermalization, an 
HIDM particle will therefore undergo a random walk that takes it a radial distance 
$\lambda_* \sim \lambda (3{\bar f}_{\rm KE})^{-1/2}$ from where it started.  In the Earth's crust, \bzzzz{the mean atomic mass is $21.5\,m_{\rm p}$,
and the mean free path is $\lambda = 35.9\,(\sigma^{\rm cr}_{\rm 300\,K}/10^{-24} \rm cm^2)^{-1} (\rho_{\rm cr} / \rm g \, cm^{-3})^{-1}$
~cm, where $\sigma^{\rm cr}_{\rm 300\,K}$ is the mean cross-section for the scattering of HIDM by atoms in the crust  
and $\rho_{\rm cr}$ is the density.  For an assumed density of $2.7\, \rm g\, {\rm cm}^{-3}$ (Dziewonski \& Anderson 1981), we obtain 
$\lambda = 13.3\,(\sigma^{\rm cr}_{\rm 300\,K}/10^{-24} \rm cm^2)^{-1}$~cm.}
The length scale on which the temperature changes in the crust, $-(d\ln T/dr)^{-1}$, is $\sim 10$~km (Fridleifsson et al. 2008).  For $m_{\rm DM} = 2$, we obtain
${\bar f}_{\rm KE} = 0.21$, and find that the criterion $-(d\ln T/dr)^{-1} \gg \lambda_*$ is satisfied for 
$\sigma^{\rm cr}_{\rm 300 K} \gg 1.7 \times 10^{-29}\rm \, cm^2$.}  \bzzzz{In the atmosphere, the mean atomic mass is $14.5\,m_{\rm p}$,
and the mean free path is $\lambda = 0.53\,(\sigma^{\rm atm}_{\rm 300\,K}/10^{-24} \rm cm^2)^{-1} (\rho_{\rm atm} / 10^{-3}\rm g \, cm^{-3})^{-1}$~km, where $\sigma^{\rm atm}_{\rm 300\,K}$ is the mean cross-section for the scattering of HIDM by atoms in the atmosphere and
$\rho_{\rm atm}$ is the atmospheric density ($1.3\,\rm g\,cm^{-3}$ at sea-level).}

For a spherically-symmetric potential, $\Phi(r)$, where $r$ is the distance from the center of the Earth, we then obtain $$ {d{\rm ln}p_{\rm DM} \over dr} = - {m_{\rm DM} \over kT\bx{(r)}} {d\Phi \over dr} = - {m_{\rm DM} g\be{(r)} \over kT\bx{(r)}} = - {Gm_{\rm DM} M_r \over r^2kT\be{(r)}}, \eqno(8)$$
where $p_{\rm DM} = n_{\rm DM}kT$ is the partial pressure of HIDM particles, $g\be{(r)} = GM_r/r^2$ is the local gravitational \be{acceleration}, \bx{$T(r)$ is the temperature}, and $M_r$ is the mass enclosed within radius $r$.  \bz{Equation (8) is exactly equivalent to the expression given by Gilliland et al.\ (1986; their equation 5) for the density distribution of DM in the Sun if the scattering cross-section is large.   A modification to the density distribution given by Gilliland et al.\ (1986) was subsequently proposed by Gould \& Raffelt (1990), but -- as discussed \bzzz{below} in Appendix A -- would appear to violate considerations of hydrostatic equilibrium} \bzz{and is inconsistent with  CC70 equation (8.1,7).}

\begin{figure}
\includegraphics[width=16 cm]{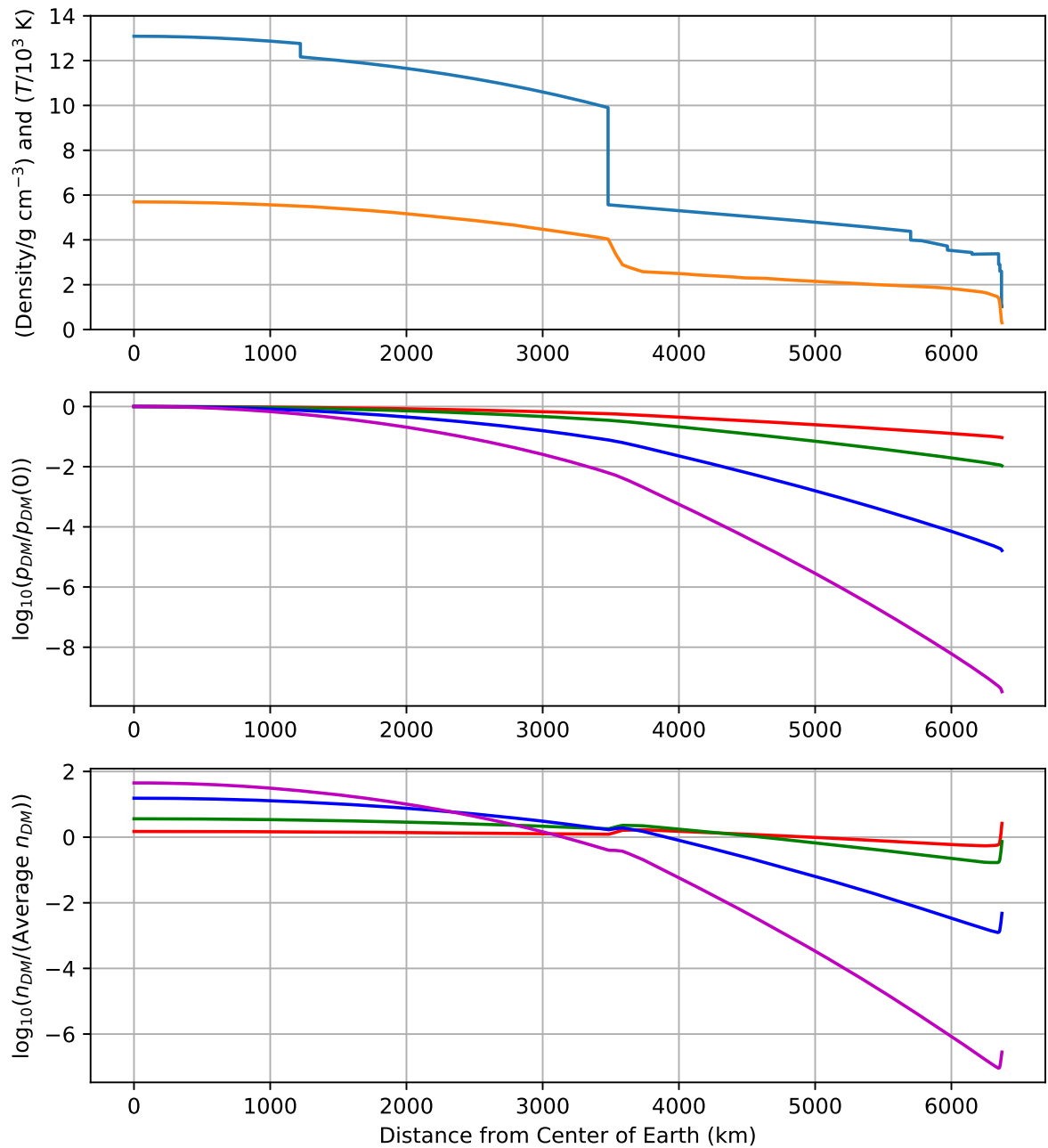}
\caption{Top: density (blue) and temperature (orange) within Earth's interior.  Middle: partial pressure of HIDM particles for $m_{\rm DM}/m_{\rm p}$ = 1 (red), 2 (green), 5 (blue) and 10 (magenta).  Bottom: volume density of HIDM particles for $m_{\rm DM}/m_{\rm p}$ = 1 (red), 2 (green), 5 (blue) and 10 (magenta).}
\end{figure}

\begin{figure}
\includegraphics[width=15 cm]{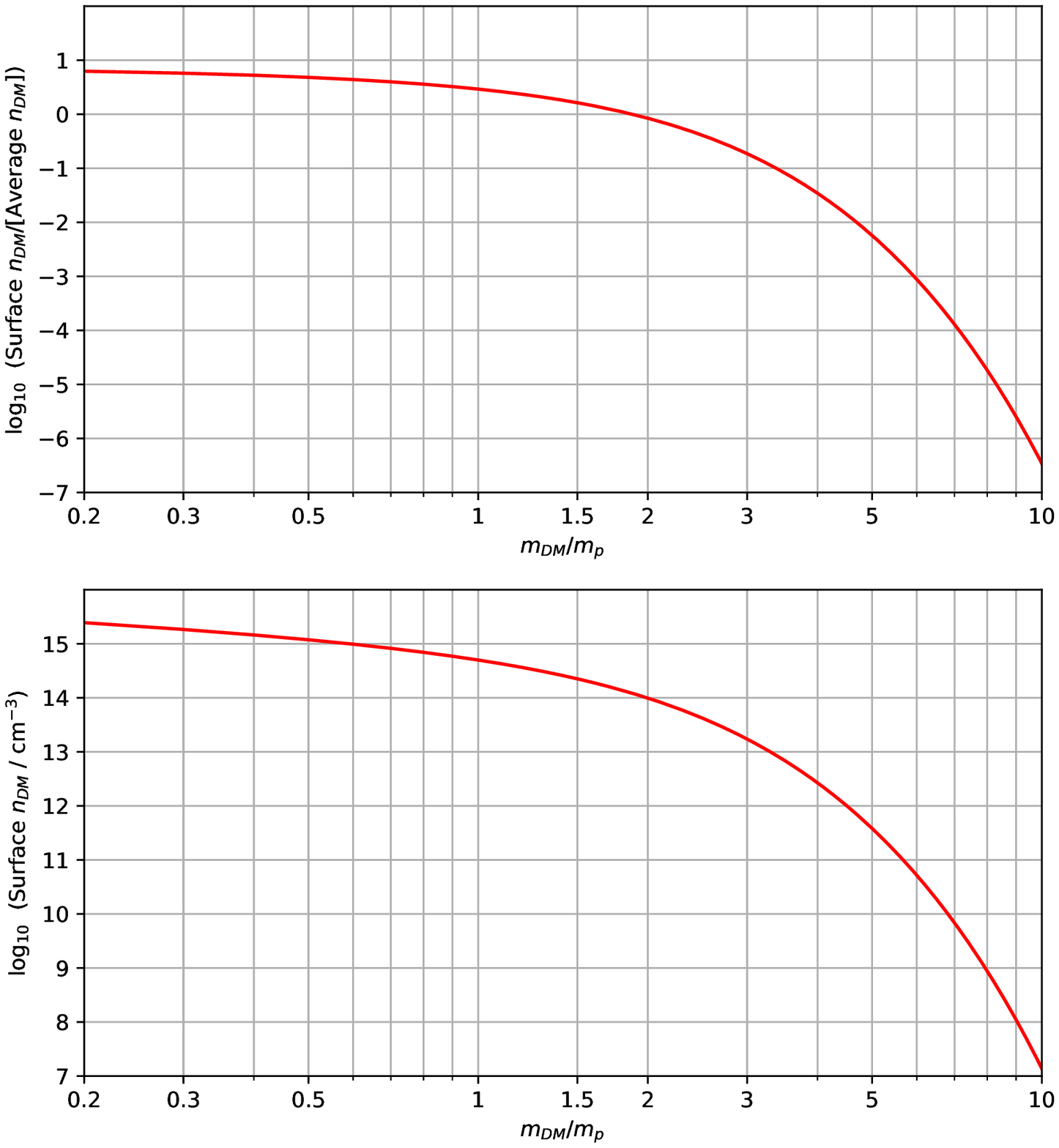}
\caption{Top: density of HIDM particles \ma{at Earth's surface}, relative to the mean density within Earth, as a function of $m_{\rm DM}/m_{p}$.   Bottom: density of HIDM particles at Earth's surface, if every particle captured is retained.  Here, an average density of 0.3~(GeV/c$^2$)~cm$^{-3}$ was assumed for the DM density at the solar circle, and a velocity of 200~km~s$^{-1}$ was assumed for the Earth's motion relative to the DM.}
\end{figure}

We integrated equation (8) \bx{ numerically} to obtain the HIDM partial pressure and density as a function of $r$,  
adopting the density profile for the Earth's interior given by Dziewonski \& Anderson (1981; the PREM model) and the temperature profile of Anzellini et al.\ (2013; their Figure S4.)
In Figure 1, we show the adopted temperature and density profiles (upper panel) and the derived HIDM pressure (middle panel), normalized relative to its central value, for four values of $m_{\rm DM}$.  \bx{For the smallest value shown here, $m_{\rm DM}=1~m_{\rm p}$, the pressure scale height is largest and the $r$ dependence is weakest (red curve).  As discussed above, we may assume here that the HIDM particles are in thermal equilibrium with the local material, because the mean free path is small compared to the length scale on which the local temperature varies.}

The bottom panel shows the corresponding HIDM particle density, normalized relative to the mean value within the Earth, $\bar n_{\rm DM}$.  
The HIDM density increases sharply to the Earth's surface, owing to the decrease in the local temperature\footnote{\bz{Analogous behavior may be observed in a household ``top freezer" refrigerator: here, the temperature in the upper freezer compartment might be 7--8 $\%$ smaller than in the lower compartment, and the density there must be correspondingly larger so that both compartments \bzz{are very close to} pressure equilibrium with each other and their surroundings (as we infer they must be because \bzz{any small fractional pressure difference would cause the 
the compartment doors to fly open or would render them unopenable, atmospheric pressure at sea-level being \bzzz{$\sim 1000$ kg} force per m$^2$).}}}, 
\bx{and can even exceed the average density for $m_{\rm DM} < 1.8 \,\be{m_{\rm p}}$.}  For $m_{\rm DM} = 2\,\be{m_{\rm p}}$, \ma{which is a natural mass for HIDM of the type proposed by Farrar (\be{2017})}, 
the \ma{density at the Earth's surface}, $n_{\rm DM}(R_\earth)$, is $0.74\,{\bar n_{\rm DM}}.$  

The upper panel of Figure 2 shows the $n_{\rm DM}(R_\earth)/ {\bar n_{\rm DM}}$ as a function of $m_{\rm DM}$, while the lower panel shows \ma{the density at the Earth's surface}, $n_{\rm DM}(R_\earth)$.  The latter was computed under the condition that every HIDM particle captured by the Earth is retained (the validity of which will be considered in section 2.4 below).  For large $m_{\rm DM}$, the density at the Earth's surface is a strongly decreasing function of $m_{\rm DM}$, primarily because the scale height of the DM particles within the Earth is inversely proportional to $m_{\rm DM}$.  Other factors affecting the \be{number} density at the Earth's surface are the number density of HIDM particles in the Galactic plane, which -- for a given mass density, $\rho_{\rm DM}$ -- is inversely proportional to $m_{\rm DM}$ \be{,} and the capture fraction, $f_{\rm cap}$, which is an increasing function of $m_{\rm DM}$ for $m_{\rm DM} \simlt 15 \, m_{\rm p}.$

\subsection{Atmospheric loss}

HIDM particles can evaporate from the Earth's atmosphere if a sufficient fraction present at or above last scattering surface (LSS) has a velocity larger than the escape velocity, $v_{\rm es} = 11.2 \, \rm km \, s^{-1}$.  In this subsection, we present an approximate treatment of this process with the purpose of obtaining an upper limit on the loss rate, as a function of $m_{\rm DM}$ and scattering cross-section.  A more precise treatment would entail a Monte-Carlo simulation that is beyond the scope of the present study.

\subsubsection{Jeans escape from the LSS}

In an isothermal atmosphere, the flux of escaping particles, $F_{\rm es}$ may be approximated by the Jeans escape formula (Jeans 1904):
$$F_{\rm es}({\rm Jeans}) = {n_{\rm LSS} v_{\rm LSS} \over 2 \pi^{1/2}} \biggl(1 + {v_{\rm es}^2 \over v_{\rm LSS}^2}\biggr)\, 
\exp(-v_{\rm es}^2/v_{\rm LSS}^2) , \eqno(9)$$
where $n_{\rm LSS}$ is the particle density, $v_{\rm LSS} = (2kT_{\rm LSS}/m_{\rm DM})^{1/2},$ and $T_{\rm LSS}$ is the temperature, each evaluated at the LSS.  As discussed by Gross (1974), the Jeans escape treatment yields a slight overestimate of the loss-rate for an isothermal atmosphere because it assumes that the tail of the particle velocity distribution is replenished instantaneously as particles escape.  

The location of the LSS depends upon the cross-section for scattering by nuclei in the atmosphere or crust.   If the LSS lies in Earth's atmosphere, the relevant cross-section is
$$\sigma_{11}^{\rm atm} = {\sum_{\rm A} n_{\rm A} \sigma_{\rm 11}^{\rm A} \over \sum_{\rm A} n_{\rm A}} , \eqno(10)$$ where $n_{\rm A}$ is the number density of element A and the sum is taken over all elements present in the atmosphere.  The escape of HIDM is strongly dominated by particles with velocities just above $v_{\rm es} = 11.2 \, \rm km\,s^{-1}$; thus, in the circumstance that the cross-sections are velocity-dependent, the values for a collision velocity of $v_{\rm es}$ are the appropriate ones for use in equation (10).  The LSS is located at the $\tau=1$ surface, at height $z_{\rm LSS}$, for which 
$$ \sigma_{11}^{\rm atm} \int_{z_{\rm LSS}}^\infty \sum_{\rm A} n_{\rm A}(z) dz = 1. \eqno(11)$$
For $\sigma_{11}^{\rm atm} \simlt \gr{2.5} \times 10^{-26} \, \rm cm^2$, the optical depth, $\tau$, at the surface of the Earth's surface falls below unity, and the LSS drops into the crust.  In this case, the relevant cross-section is $\sigma_{11}^{\rm cr}$, which is defined in manner exactly analogous to $\sigma_{11}^{\rm atm}$ (the only difference being that the values of $n_{\rm A}$ in the right-hand-side of equation (10) reflect the elemental composition of the crust, not the atmosphere.)  \ma{We introduce the quantity $\sigma_{11}^{\rm es}$, which is the cross-section that determines the location of the LSS and thereby the escaping flux, $F_{\rm es}$. For $\sigma_{11}^{\rm atm} \gr{ \simgt 2.5} \times 10^{-26} \, \rm cm^2$, we adopt $\sigma_{11}^{\rm es} = \sigma_{11}^{\rm atm}$.  Otherwise, $\sigma_{11}^{\rm es} = \sigma_{11}^{\rm cr}$.}

\begin{figure}
\includegraphics[width=15 cm]{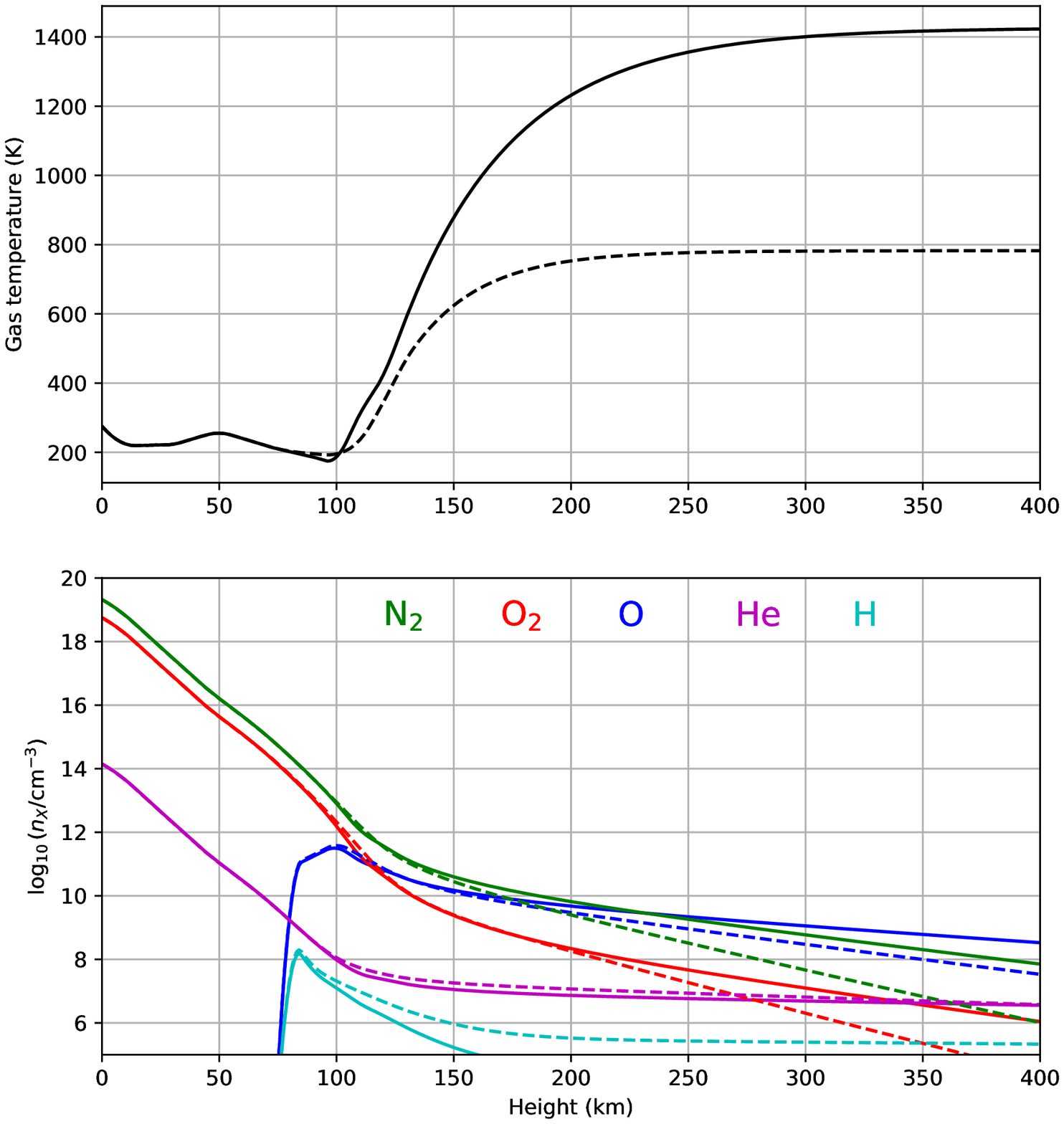}
\caption{Example temperature and density profiles adopted for the atmosphere.  Solid curves: NRLMSIS-00 results for 2001 Oct 01, a time of high solar activity (F10.7 = 263.3, F10.7A = 216.0, $A_p = 48.0$).  Dashed curves: NRLMSIS-00 results for 2007 Oct 01, a time of low solar activity (F10.7 = 65.1, F10.7A = 68.1, $A_p = 7.3$).  All results were obtained for latitude = +45 deg, longitude = 0, at 12:00 UT.}
\end{figure}

\begin{figure}
\includegraphics[width=17 cm]{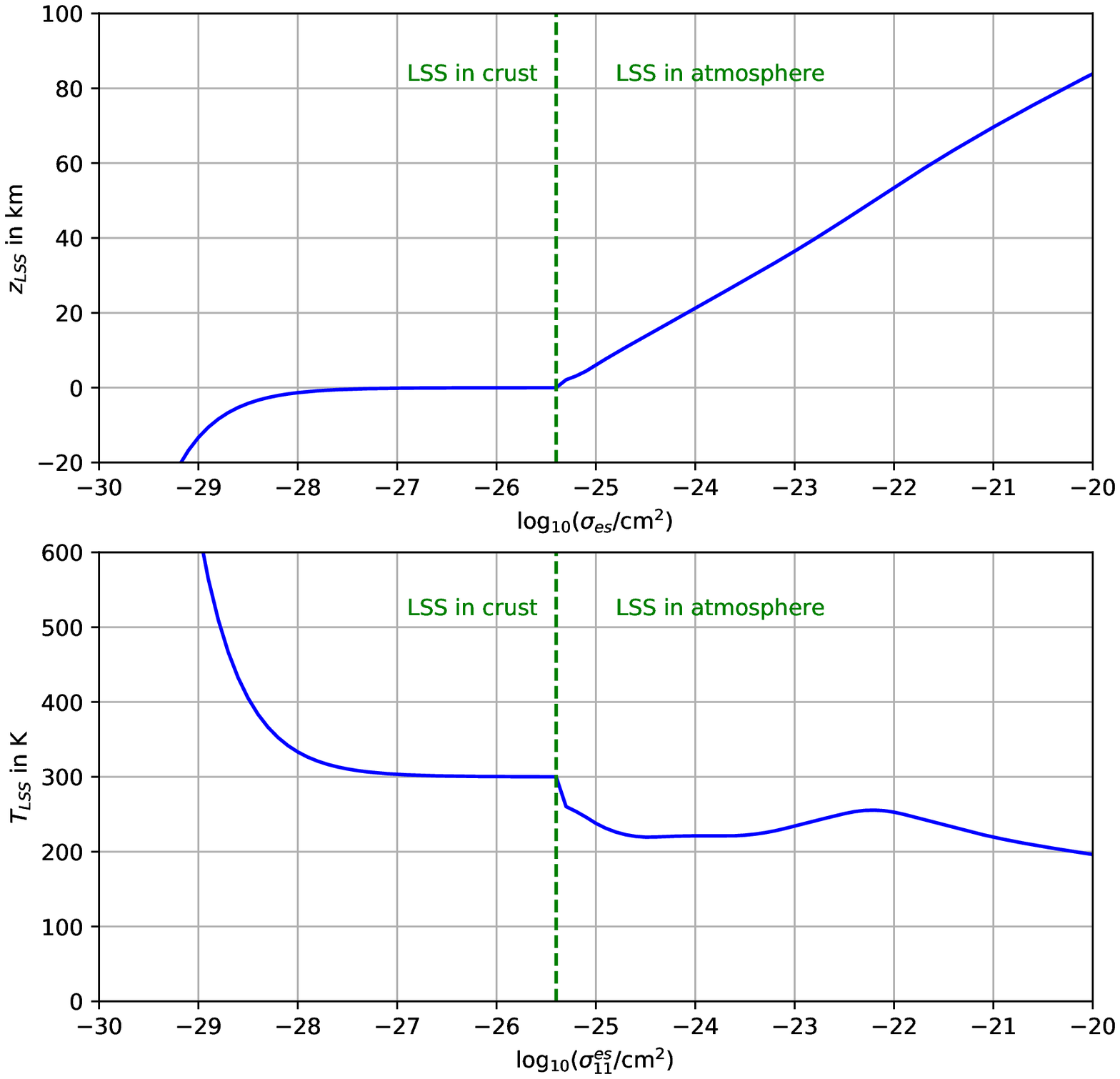}
\caption{Upper panel: location of the last scattering surface, as a function of $\sigma_{11}^{\rm es}$.  The LSS is located in the Earth's atmosphere to the right of the vertical green dashed line, and in the crust to the left of that line.  Bottom panel: temperature at the LSS.}
\end{figure}

Because the escape rate depends strongly on the gas temperature, calculating the escape rate requires an accurate knowledge of the atmospheric temperature profile.  In our treatment of atmospheric loss, we have made use of the NRLMSIS-00 model (Picone et al.\ 2002) for the density, temperature, and composition of the atmosphere at heights, $z$, up to 1000~km.  This well-tested model, which was motivated primarily by the need to model atmospheric drag on satellites in low Earth orbit, provides results as a function of latitude, longitude, height, day-of-year, hour-of-day, and solar activity, the latter being characterized by the 10.7 cm solar radio flux, F10.7, its 81 day average, F10.7A, and the $A_{\rm p}$ index of geomagnetic activity.  Two example temperature and density profiles are shown in Figure 3, one for a low level of solar activity (dashed curves) and one for a high level of solar activity.  The temperature profile below $z \sim 100$~km is not strongly dependent on the solar activity level, and the temperature lies in the range $\sim 200 - 300$~K.  Within the thermosphere ($z \simgt 100$~km), however, the temperature rises rapidly and is strongly affected by solar activity.  For each of these profiles, the location and temperature of the LSS are shown in Figure 4 as a function of $\sigma_{11}^{\rm es}$.

Based on the temperature and densities plotted in Figure 3, we have evaluated the Jeans escape formula as a function of $m_{\rm DM}$ and $\sigma_{11}^{\rm es}.$   The results are conveniently represented as a fractional loss rate,
$$f_{\rm loss}({\rm Jeans}) = {3 F_{\rm es}({\rm Jeans}) \over R_\earth {\bar n}_{\be{\rm DM}}},\eqno(12)$$
which is plotted in Figure 5. 
Because the LSS is below 100~km for the entire parameter space under consideration here (Figure 4), there is no dependence on the level of solar activity.  As expected, $f_{\rm loss}$ is a monotonically decreasing function of $m_{\rm DM}$.  For $\sigma_{11}^{\rm es} \simgt \gr{2.5} \times 10^{-26}\rm \, cm^{2}$, the non-monotonic dependence of $f_{\rm loss}$ on $\sigma_{11}^{\rm es}$ reflects the temperature profile in the atmosphere.  As $\sigma_{11}^{\rm es}$ falls below $\sim \gr{2.5} \times 10^{-26}\rm cm^{2},$ the LSS moves down into the crust, where the temperature increases with an assumed gradient $\sim 25$~K per km of depth (\gr{Fridleifsson et al.\ 2008}).  This leads to a rapid increase in the fractional loss rate for $\sigma_{11}^{\rm es} \simlt 10^{-28} \rm \, cm^{2}$.  The red dashed line represents $1/t_\earth$.  When the fractional loss rate $\simgt 1/t_\earth$, then the effects of HIDM loss are important.

\begin{figure}
\includegraphics[width=15 cm]{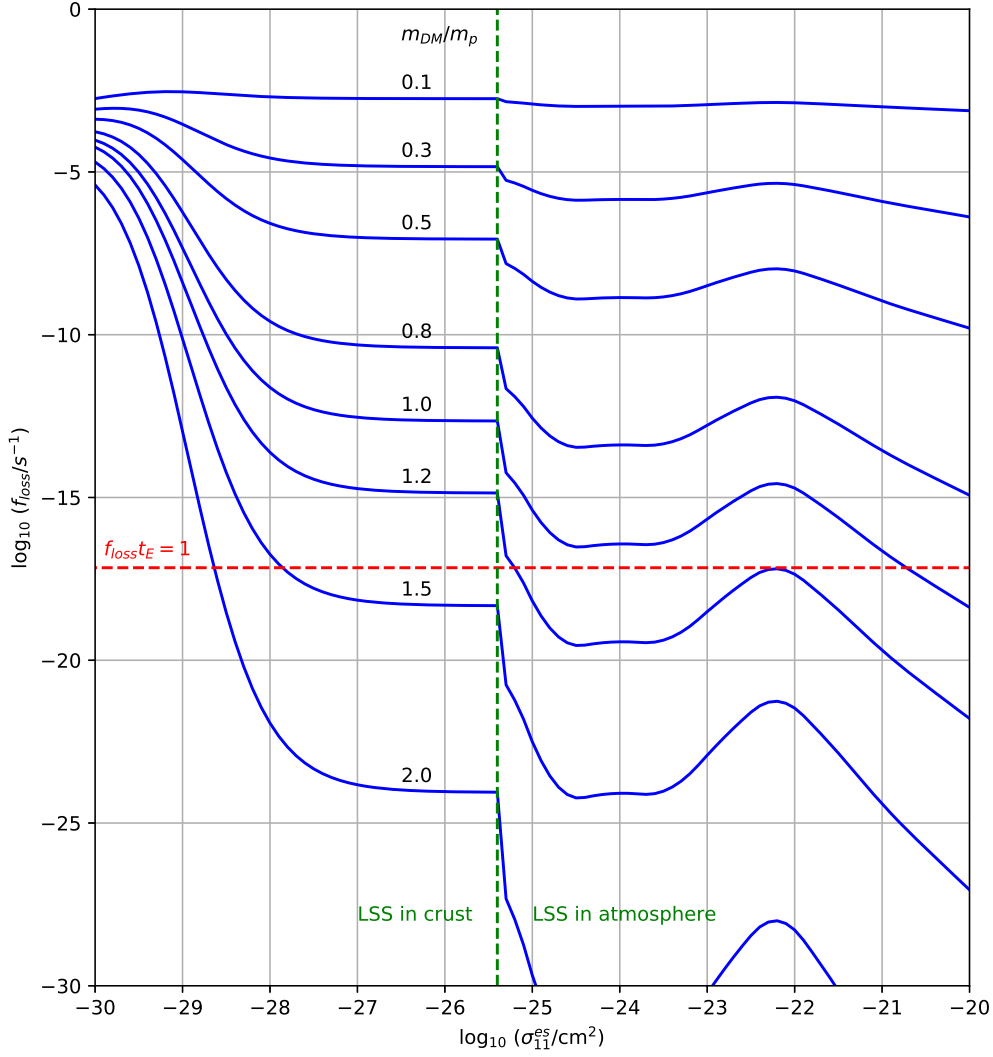}
\caption{Fractional loss rates given by the Jeans escape formula, as a function of $\sigma_{11}^{\rm es}$ and for several values of $m_{\rm DM}/m_p$.  Atmospheric loss is important above the horizontal red dashed line, $f_{\rm loss}({\rm Jeans}) = 1/t_\earth$.  The LSS is located in the Earth's atmosphere to the right of the vertical green dashed line, and in the crust to the left of that line.}
\end{figure}
 
\subsubsection{Thermospheric escape from above the LSS}

In the Jeans treatment of evaporation from planetary atmospheres, the loss rate is determined solely by the temperature and density at the LSS (equation 9).  In the case considered here, however, the atmospheric temperature increases rapidly above the LSS, within the thermosphere at $z = 100 -200$~km (Figure 4).  Because the loss rate depends very strongly on the temperature, this rapid increase raises the question of whether the loss rate might be enhanced significantly above the value given by equation (9).  Such an enhancement might result from collisions between hot atmospheric molecules in the thermosphere and HIDM, but its importance is mitigated by two effects: (1) the gas density declines rapidly with $z$; and (2) a single collision between an HIDM and a hot thermospheric gas molecule only transfers a fraction $f_{\rm KE}$ of the hot molecules' energy. 
In Appendix B, we present an analysis of the effects of thermospheric escape.  Because of the second consideration above, thermospheric escape is controlled by collisions with the minor atmospheric constituents H and He; when $m_{\rm DM}$ lies in the range when thermospheric escape is important, $f_{\rm KE}$ is largest for these lightest atmospheric constituents.  Our analysis is complicated by the fact that the thermospheric temperature depends strongly on solar activity: this results in large variations during the 11-year solar cycle, which we account for with the NRLMSIS-00 model.  \ma{For the range of parameters under present consideration, the result of our analysis is that thermospheric escape is always negligible \bzzz{relative to $1/t_\earth$ or to the standard Jeans loss rate} if (1) $\sigma^{\rm He}_{11}$ and $\sigma^{\rm H}_{11}$ are both smaller than  $10^{-24}\, \rm cm^2$; or (2) if $m_{\rm DM}$ is smaller than 0.7~$m_{\rm p}$.  
As we shall see in Section 3.3 below, upper limits on $\sigma^{\rm He}_{300 K}$ and $\sigma^{\rm H}_{300 K}$  may be obtained from the low vaporization rates for liquid He and H$_2$ achievable in well-insulated dewars.  Under the assumption that the cross-sections at $v_{\rm es}$ are no larger than those at 300~K \be{and are no larger than $10^{-20}\, \rm cm^2$}, the combination of these constraints implies that thermospheric escape is \gr{also negligible for $m_{\rm DM} \simgt 1.2 \,m_{\rm p}$.}}

An important caveat pertains to that result, however.  Our model for the time-averaged loss rate is severely limited by the short historical record of solar activity.  We have only computed loss-rates over the past $\sim 50$ years, for which measurements of the F10.7A and  $A_{\rm p}$  activity indicators are available.   Sunspot observations, which are available back into the 17th century, suggest that solar activity varies on multiple timescales that can be much longer than the 11-year solar cycle.  Most notably, sunspots were exceedingly rare during the Maunder mimimum 
(1645 -- 1715), implying an unusually low level of solar activity, and the particle loss rate was presumably very low as a result.  Equally, we cannot exclude the possibility that extended periods of very high solar thermospheric loss rate have occurred prior to the 17th century.  Thus, because of the exponential dependence of the loss rate on the level of solar activity, the average loss rate over geological timescales could differ greatly from the average we determined for the last 50 years.

\subsection{Density of HIDM at the surface of the Earth}

If HIDM particles are being captured a rate $f_{\rm cap} \rho_{\rm DM} v_\earth \pi R_\earth^2 m_{\rm DM}^{-1}$, and are being lost at a fractional rate $f_{\rm loss}$, the number within the Earth varies according to 
$${dN \over dt} = {f_{\rm cap} \rho_{\rm S} v_\earth \pi R_\earth^2 \over m_{\rm DM}} - N f_{\rm loss}. \eqno(13)$$
If the capture and fractional loss rates have been constant over the history of the Earth, the current number of HIDM particles is
$$N(t_\earth) = {f_{\rm cap} \rho_{\rm DM} v_\earth \pi R_\earth^2 \bzzz{t_\earth} \over m_{\rm DM}} \times
{1 - \exp(-f_{\rm loss} t_\earth) \over f_{\rm loss} t_\earth} \eqno(14)$$ 
The factor $[1 - \exp(-f_{\rm loss} t_\earth)] / [f_{\rm loss} t_\earth]$ represents the correction that is needed to account for atmospheric loss.  It is equal to unity in the limit $f_{\rm loss} \ll 1/t_\earth$, and tends to $[f_{\rm loss} t_\earth]^{-1}$ in the limit $f_{\rm loss} \gg 1/t_\earth.$

In Figure 6, we plot contours of the number density at the Earth's surface, $n_{\rm DM}(R_\earth),$
in the m$_{\rm DM}- \sigma_{11}^{\rm es}$ plane.  Unlike in Figure 2, these results include the effects of atmospheric loss.  
Particle densities in excess of $ 10^{14}\,\rm cm^{-3}$ can be achieved for $m_{\rm DM}$ in the range $1 - 2 \, m_p.$  Atmospheric escape becomes increasingly important at very small $\sigma_{11}^{\rm es}$, because the LSS falls deep into the crust or upper mantle where the temperature is higher.  The particle density at Earth's surface also drops rapidly for $m_{\rm DM} > 3\,m_{\rm p}$, because the particles become increasingly concentrated toward the center of the Earth.  The results shown in Figure 6 apply under conditions (see Section 2.4.2 above) where thermospheric loss can be neglected,

\begin{figure}
\includegraphics[width=17 cm]{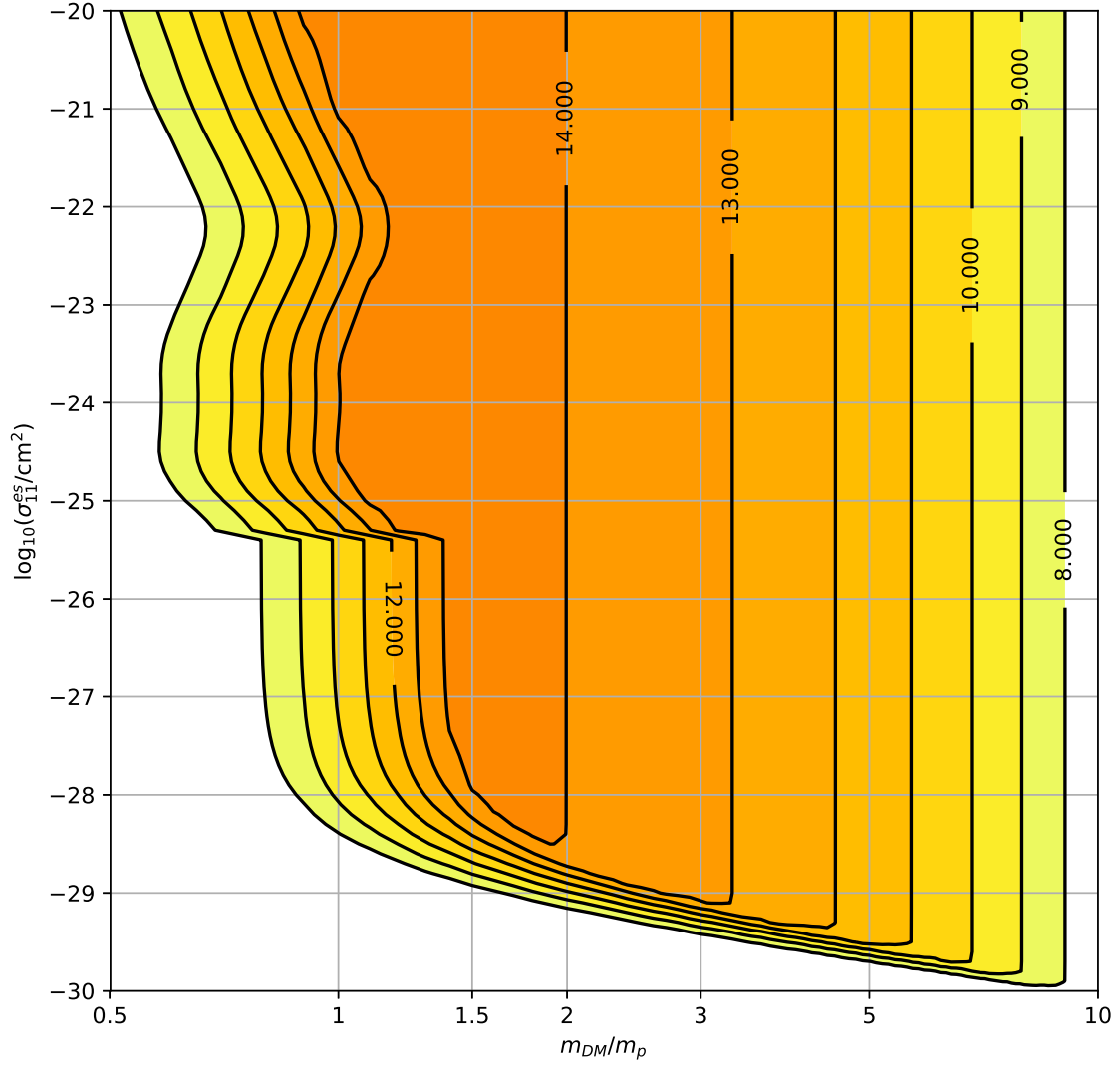}
\caption{Number density of HIDM at the Earth's surface, $n_{\be{\rm DM}}(R_\earth),$
in the m$_{\rm DM}- \sigma_{11}^{\rm es}$ plane. Contours are labeled by log$_{10} (n_{\rm DM} / \rm cm^{-3})$.}
\end{figure}

\section{Limits on the HIDM density at the surface of the Earth}

Figure 6 suggests that the density of captured HIDM particles at the Earth's surface could exceed the interstellar particle density by $\sim 15$ orders of magnitude for particle parameters $(m_{\rm DM}, \sigma_{11}^{\rm es})$ within the range considered here.  These thermalized particles would have typical kinetic energies $kT \sim 0.025$~eV,  
significantly below the threshold for detection in current DM searches.  Nonetheless, their presence could have detectable effects in experiments that are not specifically designed for DM detection.  In this section, we discuss the constraints on the particle parameters that are placed by measurements of the beam lifetime in the Large Hadron collider (LHC), the orbital decay of spacecraft in LEO, the vaporization rate of liquid helium (LHe) \bzzz{and other cryogens}, and the temperature gradient in the Earth's crust.  Except where otherwise stated, the limits we derive apply under conditions where  the Jeans escape formula (eqn.\ 9) is applicable, \be{and the thermospheric loss process discussed in Appendix B can be neglected}.

\subsection{The LHC beam lifetime}

Within a significant region of parameter space, $n_{\rm DM}(R_\earth)$ significantly exceeds the gas density in the LHC beam pipe, which is evacuated to high vacuum ($\sim 10^{-7}$~Pa, equivalent to a gas density $\sim 10^9\, \rm cm^{-3}$) in order to achieve a long mean free path for the relativistic protons.  \gr{Inelastic collisions with particles within the beam pipe lead to a reduction of 
the beam intensity on a timescale that can be as large as 100~hr (Lamont \& Johnson 2014), requiring a mean free path $\lambda_{\rm LHC} > 1.1 \times 10^{16} \, \rm cm$ for inelastic scattering of protons traveling close to the speed of light.  Elastic collisions \be{are} not important here, because the magnetic multipoles correct for momentum transfer that is unaccompanied by significant energy loss.}\footnote{\gr{https://www.lhc-closer.es/taking$\_$a$\_$closer$\_$look$\_$at$\_$lhc/0.beam$\_$lifetime, downloaded on 2018~March~25}} 
The mean free path for \gr{inelastic} collisions with HIDM is $\bx{[n_{\rm DM}(R_\earth)\sigma_{\rm 6.5\,TeV}^{\rm p, inel}]^{-1}}$, where $\sigma_{\rm 6.5\,TeV}^{\rm p, inel}$ is the cross-section for the \gr{inelastic} scattering of 6.5~TeV protons by stationary HIDM.

An LHC beam lifetime of 100 hr places an upper limit on the high energy inelastic scattering cross-section 
$\sigma_{\rm 6.5\,TeV}^{\rm p, inel} < 9 \times 10^{-17} \bx{[n_{\rm DM} (R_\earth)/\rm cm^{-3}]^{-1} {\rm cm^2}}.$
In Figure 7, this constraint is plotted in the $m_{\rm DM} - \sigma_{\rm 6.5\,TeV}^{\rm p, inel}$ plane.  \bx{  Here, we adopted the values of $n_{\rm DM} (R_\earth)$ plotted in Figure 6; for the cross-sections under present consideration, the pumping of atmospheric gases out of the beam pipe does not alter the density of HIDM, because the latter are constantly diffusing through the pipe walls into the evacuated region.} 
Where HIDM escape is significant, $n_{\rm DM}(R_\earth)$ depends on $\sigma_{11}^{\rm es}$ as well as $m_{\rm DM}$.  Thus, the limits obtained on $\sigma_{\rm 6.5\,TeV}^{\rm p, inel}$ depend on what is assumed for $\sigma_{11}^{\rm es}$.  In Figure 7, we have plotted results applying for five values of $\sigma_{11}^{\rm es}$ for which the LSS lies in the crust (\bzzz{$10^{-29.0}$, $10^{-28.8}$}. $10^{-28.5}$, $10^{-28}$, and $10^{-27} \, \rm cm^2$), and two values for which the LSS lies in the atmosphere ($10^{-22}$ and $10^{-20}\,\rm cm^2$).  

\begin{figure}
\includegraphics[width=14 cm]{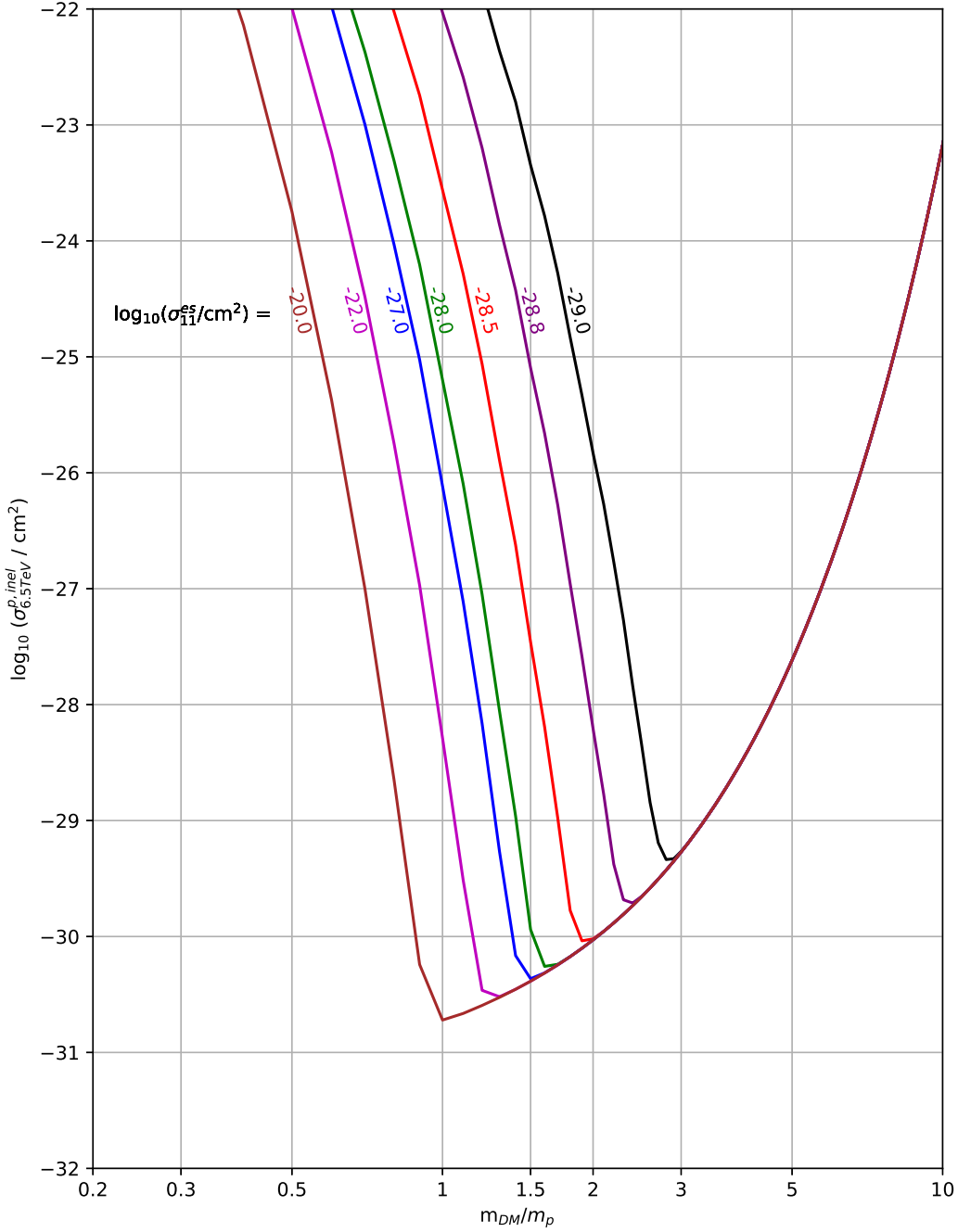}
\caption{Upper limits on $\sigma_{\rm 6.5\,TeV}^{\rm p, inel}$ implied by an LHC beam lifetime of 100 hr.  Results are shown for \bzzz{five} values of $\sigma_{11}^{\rm es}$ for which the LSS lies in the crust (\bzzz{$10^{-29.0}$, $10^{-28.8}$}, $10^{-28.5}$, $10^{-28}$, and $10^{-27}\,\rm cm^2$), and two values for which the LSS lies in the atmosphere ($10^{-22}$ and $10^{-20}\,\rm cm^2$).  The curves are labeled with log$_{10}(\sigma_{11}^{\rm es})$.}
\end{figure}


\subsection{The orbital decay of spacecraft in low Earth orbit}

Spacecraft orbiting within the Earth's thermosphere can experience a significant drag force, $F_{\rm drag}$, which results from collisions with atmospheric molecules and leads to orbital decay.  Within a significant region of the parameter space considered here, the mass density of HIDM can greatly exceed the mass density of atmospheric molecules.  \bzz{Above the last scattering surface (LSS), the HIDM partial pressure continues to decline according to equation (8), but with $T_{\rm LSS}$ replacing $T(r)$ in that equation.  Thus, in the limit $z_{\rm LSS} \le h \ll R_\earth$, the particle density at altitude $h$ is $n_{\rm DM}(h) = n_{\rm DM} (z_{\rm LSS}) \exp[-m_{\rm DM}g(h-z_{\rm LSS})/kT_{\rm LSS}]$, where the gravitational acceleration, $g$, may be approximated by its value at the Earth's 
surface.}\footnote{\bzz{This result for $n(h)$ may be demonstrated by considering the upward flux of particles of velocity $v_z$ leaving the LSS, $F_{vz} dv_z = n_{\rm LSS} (m_{\rm DM}/[2 \pi kT_{\rm LSS}])^{1/2} \exp(-m_{\rm DM}v_z^2/[2kT_0])dv_z$. Here, we make an approximation in assuming that HIDM \bzzz{particles} at the LSS have a Maxwell-Boltzmann velocity distribution function. 
Such particles reach a maximum altitude, $h_{\rm max} = z_{\rm LSS} + v_z^2/2g$, and remain above altitude $h$ for a time period $t_h = 2[(2h_{\rm max} - 2h)/g]^{1/2}$.  Integrating $F_{\rm vz}t_h dv_z$ over all $z$-velocities for which $h_{\rm max} > h$, we find that the column density of HIDM particles {\it above} altitude $h$ is 
$N(h) = n_{\rm DM}(z_{\rm LSS})(kT_{\rm LSS}/m_{\rm DM}) \exp(-m_{\rm DM}g(h-z_{\rm LSS})/[kT_{\rm LSS}])$. 
The HIDM density {\it at} altitude, $h$, is then given by  $-dN(h)/dz$, \bzzz{which yields} the expression obtained above from equation (8) with the use of $T_{\rm LSS}$ in place of $T(r)$.}}   
In Figure 8, we show the expected number density of HIDM at a height of 600~km above Earth's surface,
as a function of $m_{\rm DM}/m_{\rm p}$ and $\sigma_{11}^{\rm es}$.  For $m_{\rm DM} = 2\,m_{p}$, and $\sigma_{11}^{\rm cr}$ in the range $10^{-28}$ to $\gr{2.5} \times 10^{-26}\,\rm cm^2$, the HIDM number density \be{at 600~km} is \bzzz{$8.5 \times 10^{11} \,\rm cm^{-3}$,} corresponding to a mass density of $2.8 \times 10^{-12} \rm g\,cm^{-3}$.
For comparison, the mass densities of atmospheric gases at 600~km implied by Figure 3 are $3.3 \times 10^{-15} \rm g\,cm^{-3}$ and  $5.9 \times 10^{-17} \rm g\,cm^{-3}$, respectively, for the low and high solar activity cases.   In the regime where the HIDM mass density greatly exceeds these atmospheric values, 
the orbital decay of spacecraft in low Earth orbit would be greatly accelerated unless the scattering cross-sections are low enough that most HIDM particles pass through the spacecraft.

\begin{figure}
\includegraphics[width=15 cm]{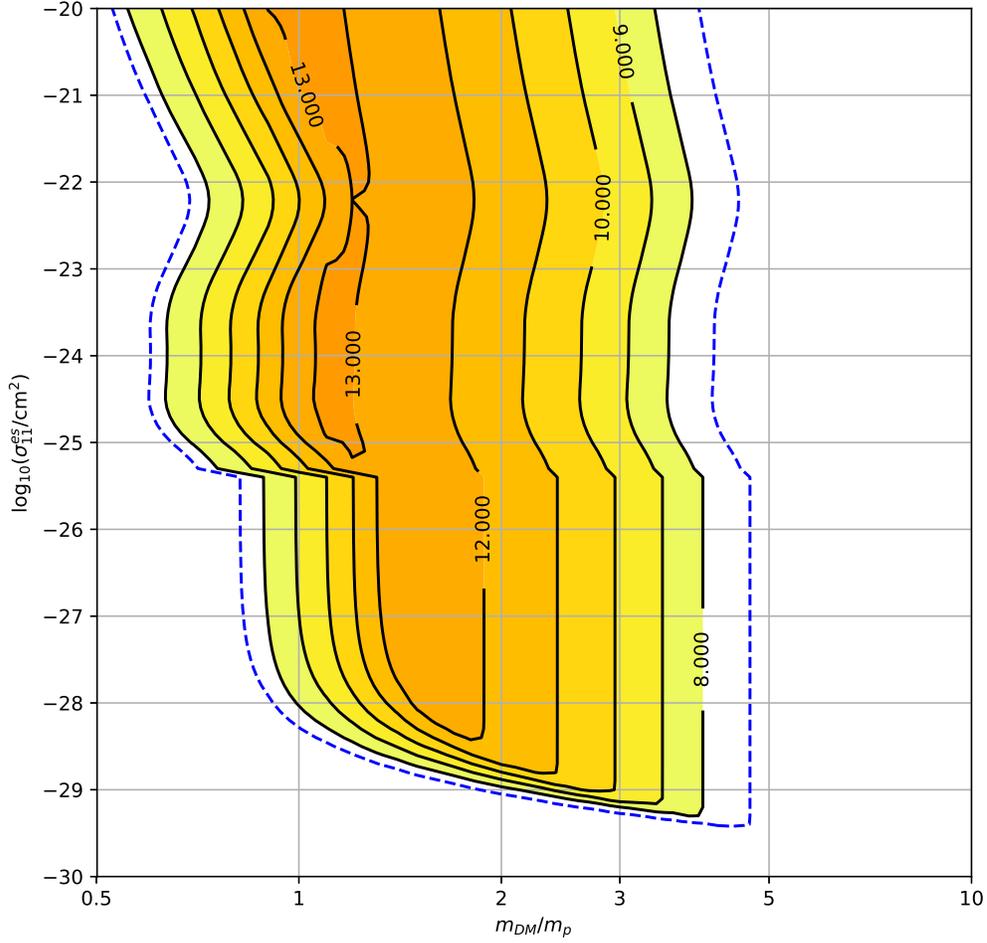}
\caption{Number density of HIDM at an altitude of 600~km.  Contours are labeled by log$_{10} (n_{\rm DM} / \rm cm^{-3})$.  The dashed blue contour encloses the region where the orbital decay rate of HST would exceed the observed value if HST were opaque to HIDM (see the text).}
\end{figure}

In the limit of small \ma{scattering cross-section}, \gr{such that the probability of more than one scattering is negligible}, the drag force due to HIDM is
$$F_{\rm drag} = n_{\rm DM}(R_{\rm orb}) \bar{\mu} v_{\rm orb}^2 (M_{\rm sc}/{\bar m}_{\rm A}) \sigma_8^{\rm sc}\eqno(15)$$
where $M_{\rm sc}$ is the spacecraft mass, $R_{\rm orb}$ is the radius of the orbit, 
$v_{\rm orb} = 7.9 (R_{\rm orb}/R_\earth)^{-1/2}\,\rm km \, s^{-1}$ is the orbital velocity, \ma{ ${\bar m}_{\rm A}$ is the mean atomic mass for the material constituting the spacecraft}, \bzzz{$\bar{\mu}$ is the mean reduced mass}, and
$\sigma^{\rm sc}_8$ is the average cross-section per \ma{nucleus} at a collision 
velocity of $v_{\rm orb} \sim \rm 8\, km\,s^{-1}$.  \be{Henceforth, we assume that the HIDM mass is small compared to the mass of typical nuclei in the spacecraft, so the scattered HIDM have an average momentum of zero in the spacecraft frame, and $\bar{\mu}$ may be replaced by $m_{\rm DM}$ in equation (15).}  The quantity $M/{\bar m}_{\rm A}$ is simply the number of \ma{nuclei} within the spacecraft.

The total energy, $E_{\rm tot}$, of the orbiting spacecraft decreases at a rate 
$${dE_{\rm tot} \over dt} 
= -F_{\rm drag} v_{\rm orb}
= -n_{\rm DM}(R_{\rm orb}) \biggl( {m_{\rm DM} \over {\bar m}_{\rm A}}\biggr) M_{\rm sc} v_{\rm orb}^3   \sigma^{\rm sc}_8 \eqno(16)$$ 
Noting now that $E_{\rm tot} = -E_{\rm kin} = -\onehalf M_{\rm sc}v_{\rm orb}^2$, in accord with the virial theorem, and that $E_{\rm tot}$ is inversely proportional to $R_{\rm orb}$,
we may write the orbital decay rate in the form
$${dR_{\rm orb} \over dt} =  - {R_{\rm orb} \over E_{\rm tot}} {dE_{\rm tot}\over dt} = - \be{2} n_{\rm DM}(R_{\rm orb}) \biggl( {m_{\rm DM} \over \ma{{\bar m}_{\rm A}}}\biggr) 
R_{\rm orb} v_{\rm orb} \sigma^{\rm sc}_8  $$
$$=\ma{-11.7} \biggl({n_{\rm DM}(R_{\rm orb}) \over 10^{12} {\rm cm}^{-3}}\biggr) \biggl({R_{\rm orb} \over R_\earth} \biggr)^{-1/2} \biggl( {m_{\rm DM} \over m_{\rm p}}\biggr) 
\biggl( \ma{\bar{m}_{\rm A} \over \ma{27}\,m_p}\biggr)^{-1}
\biggl({\sigma^{\rm sc}_8 \over \rm mb}\biggr) \,\rm km \, yr^{-1}. \eqno(17)$$
\ma{Here, we have normalized the mean atomic mass relative to the value for aluminum.}

The orbital decay rate, $dR_{\rm orb}/dt$, scales linearly with $\sigma^{\rm sc}_8$, until the spacecraft becomes opaque to HIDM.  In the limit of large $\sigma^{\rm sc}_8$, the drag force becomes independent of the cross-section:
$$F_{\rm drag} = \onehalf C_{\rm d} n_{\rm DM}(R_{\rm orb}) m_{\rm DM} \Sigma_{\rm eff} v_{\rm orb}^2,\eqno(18)$$
where $\Sigma_{\rm eff}$ is the effective cross-section presented by the spacecraft and $C_{\rm d}$ is the drag coefficient.  The orbital decay rate then becomes
$${dR_{\rm orb} \over dt} =  \bzzz{ -  C_{\rm d}} n_{\rm DM}(R_{\rm orb}) m_{\rm DM} 
R_{\rm orb} v_{\rm orb} \Sigma_{\rm eff}/M_{\rm sc}  $$
$$=- \ma{2.66} \times 10^5\, \biggl({n_{\rm DM}(R_{\rm orb}) \over 10^{12}\, {\rm cm}^{-3}}\biggr) \biggl({R_{\rm orb} \over R_\earth} \biggr)^{-1/2} \biggl( {m_{\rm DM} \over m_{\rm p}}\biggr) \biggl({\Sigma_{\rm eff}/M_{\rm sc} \over \rm cm^2 \, \rm g^{-1}}\biggr) C_{\rm d} \,\rm km \, yr^{-1}. \eqno(19)$$  \ma{Thus the actual orbital decay rate can be approximated by computing the values given by equations (17) and (19) and adopting whichever has the smaller magnitude.}

\begin{figure}
\includegraphics[width=14 cm]{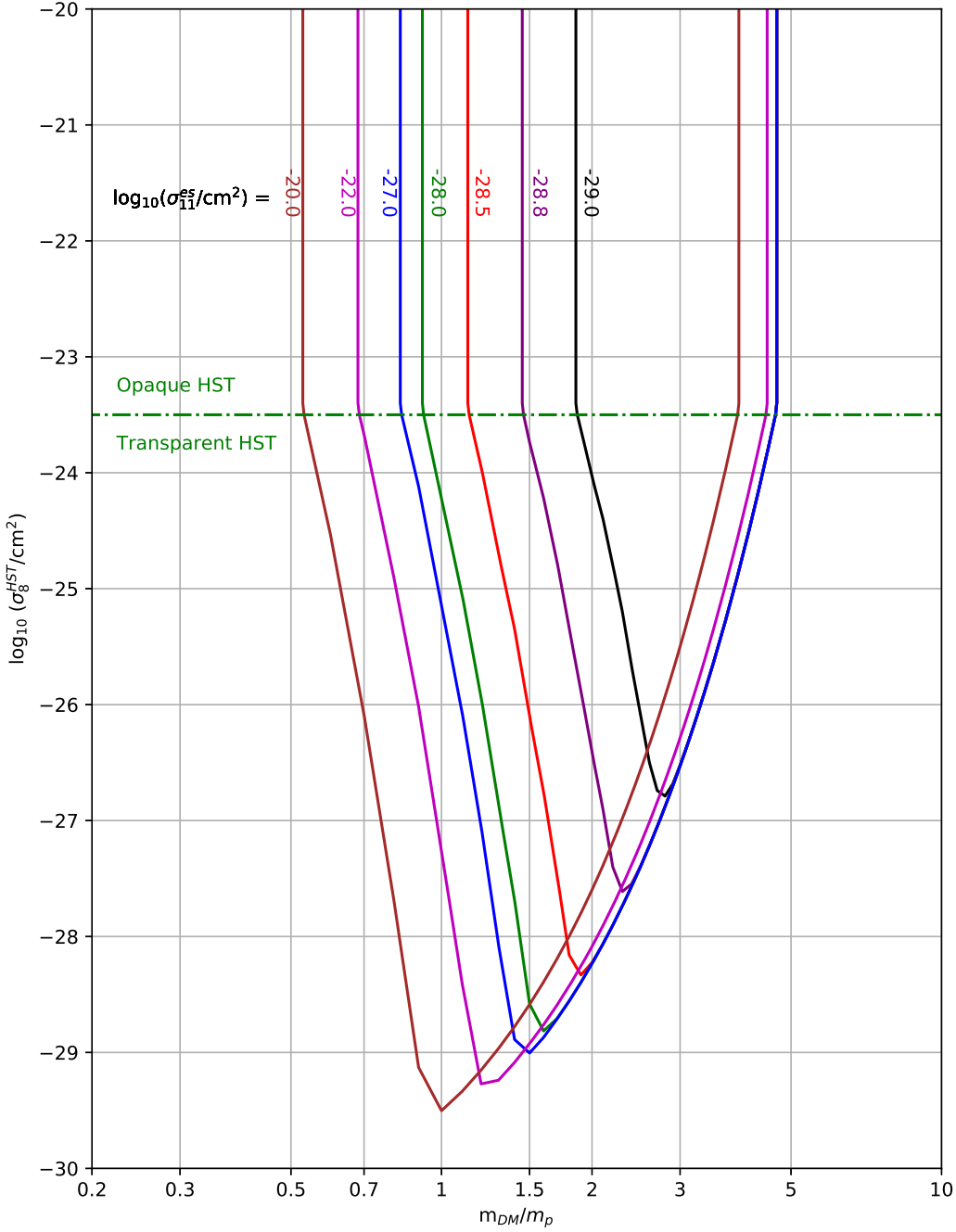}
\caption{Upper limits on $\sigma^{\rm sc}_8$, implied by 0.8 km~yr$^{-1}$ orbital decay rate for a satellite at an altitude of 600 km.  Results are shown for five values of $\sigma_{11}^{\rm es}$ for which the LSS lies in the crust (\bzzz{$10^{-29.0}$, $10^{-28.8}$}. $10^{-28.5}$, $10^{-28}$, and $10^{-27} \, \rm cm^2$), and two values for which the LSS lies in the atmosphere ($10^{-22}$ and $10^{-20}\,\rm cm^2$).  The curves are labeled with log$_{10}(\sigma_{11}^{\rm es})$.}
\end{figure}

A full discussion of the $\Sigma_{\rm eff}/M_{\rm sc}$ values, elemental composition, orbital parameters, and the observed $dR_{\rm orb} /dt$ for the fleet of spacecraft in LEO is beyond the scope of this paper.  Here, we will focus on a single example, the Hubble Space Telescope (HST), which orbits in a nearly circular orbit at an altitude of $\sim 600$~km.  \be{Elements present 
with a significant abundance in HST include Al, O, Si, C, Cu, Fe, Ti, Mg and Ni (C.~Long, personal communication).}  The orbital decay has been very well characterized and varies with solar activity.  Between servicing missions SM1 (Dec 1993) and SM2 (Feb 1997), when the solar activity was low, the average $dR_{\rm orb} / dt$ was 0.8~km~yr$^{-1}.$  In Figure 9, we show the conservative upper limit on $\sigma^{\rm HST}_8$ that is obtained with the assumption that HIDM dominated the drag on HST during this period.  The results plotted here
were obtained \ma{for an estimated ${\bar m}_{\rm A}$ of 27}, and for the spacecraft parameters\footnote{\ma{``HST Orbit Decay and Shuttle Re-boost" Fact Sheet, downloaded on 2018 March 12 from $\rm https://asd.gsfc.nasa.gov/archive/hubble/a\_pdf/news/facts/sm3b/fact\_sheet\_reboost.pdf$.}}
relevant to HST ($\Sigma_{\rm eff} = 7.0 \times 10^5 \rm cm^{2}$; $M_{\rm sc} = 1.11 \times 10^7 \rm g$; and $C_{\rm d}$=2.47).  Above the horizontal line, HST is opaque to HIDM and the orbital decay rate due to HIDM is given by equation (19).  Below the horizontal line, HST is transparent and equation (17) applies. \be{In the limit where HST is opaque to HIDM, equation (19) places an upper limit of $2 \times 10^7\, (m_{\rm DM}/m_{\rm p})^{-1}\, \rm cm^{-3}$ on $n_{\rm DM}(R_{\rm orb})$, which is indicated by the dashed blue contour in Figure 8 and excludes a large portion of the parameter space under consideration.}  

\subsection{The vaporization of liquid cryogens}
Next, we considered the heating effects of thermal HIDM particles on liquid cryogens within a well-insulated storage dewar.  \bx{  Such dewars, which are widely used to store cryogenic liquids for periods of several months without the use of active cooling, rely on multiple layers of reflective material, under vacuum, to minimize the entry of heat into the cryogenic chamber through 
conduction or radiation.}  If the HIDM mean free path is large compared to the size of the chamber, \gr{so that the dewar is transparent, and the temperature of the cryogen, $T_{\rm cry}$, is much smaller than the temperature of HIDM in the laboratory, $T_{\rm DM} \sim 300$~K, the heating rate per atom is 
$$H_{\rm \be{td}} = n_{\rm DM} (R_\earth) \langle \onehalf m_{\rm DM} v^2 \sigma_{\rm v}^{\rm A} v  \rangle_{\rm MB} {\bar f}_{\rm KE}$$ 
$$= 2\,kT_{\rm DM}\,n_{\rm DM} (R_\earth)\,{\bar v}\,\sigma_{\rm 300\,K}^{\rm A} {\bar f}_{\rm KE}, \eqno(20)$$ 
where the angled bracket denotes an average over the Maxwell-Boltzmann distribution function (eqn.\ A7), i.e.\ $\langle Y \rangle_{\rm MB} = \int_0^\infty Y(v) f_{\rm MB} dv$ for any quantity $Y$.
Here, the subscript ``\be{td}" denotes the ``transparent dewar" limit, and $\sigma_{\rm T}^{\rm A} = \langle v^3 \sigma_{\rm v}^{\rm A} \rangle_{\rm MB} / \langle v^3 \rangle_{\rm MB}$ is the appropriately-weighted mean cross-section for thermal HIDM particles at temperature $T$.}  \bu{If the cross-section has a power-law velocity-dependence of the form 
$\sigma_{\rm v}^{\rm A} = \sigma_{\rm 0}^{\rm A} (v/v_0)^{-j}$,  we find that $\sigma_{\rm T}^{\rm A} = \onehalf \Gamma(3-\onehalf j) (2 k T/\be{[m v_0^2]})^{-j/2} \sigma_{0}$.  For the case where $j=4$ (velocity-dependence for Rutherford scattering) and $v_0 = 1\,\rm km\,s^{-1}$, we obtain $\sigma_{\rm T}^{\rm A} = 0.0204 \, (T/{\rm 300 \,K})^{-2} \be{(m_{\rm DM}/m_{\rm p})^{2}} \sigma_1^{\rm A}.$} 
\gr{In the case where $T_{\rm cry}$ is not much smaller than $T_{\rm DM}$, we may estimate $H_{\rm \be{td}}$ by replacing $T_{\rm DM}$ by $(T_{\rm DM}-T_{\rm cry})$ in equation (20).}  

For a cryogenic liquid composed of atoms of mass $m_{\rm A}=Am_{\rm p}$, for which the specific latent heat of vaporization is $L_{\rm vap}$, this heating rate will result in vaporization at a fractional rate
$$ \bu{-{d\ln M_{\rm cry} \over dt}} = {H_{\rm \be{td}} \over L_{\rm vap} m_{\rm A}}$$
$$ = \bu{1.24 \times 10^{-6}} 
\biggl({T_{\rm DM} - T_{\rm cry} \over 300\,\rm{K}} \biggr)
\biggl({T_{\rm DM}  \over 300\,\rm{K}} \biggr)^{1/2} 
\biggl({m_{\rm DM}  \over m_{\rm p}} \biggr)^{-1/2}
\biggl({n_{\rm DM} (R_\earth) \over 10^{14}\,\rm cm^{-3}} \biggr)
\biggl({\sigma^{\rm A}_{\rm 300\,K} \over \rm {mb}} \biggr)
\biggl({AL_{\rm vap} \over {\rm 100 \, J \,g^{-1}}} \biggr)^{-1}{\bar f}_{\rm KE}\,{\rm s}^{-1}. \eqno(21)
$$
where mb denotes the millibarn (1 mb = $10^{-27}\,\rm cm^{2}$), \bu{and $M_{\rm cry}$ is the mass of cryogen remaining in the dewar.}
Even cross-sections in the millibarn range can lead to significant vaporization.  As an example, let us consider a case in which liquid He (LHe) is heated by HIDM with a mass $m_{\rm DM} = 2\,m_{\rm p}$ and a density $n_{\rm DM} (R_\earth) = 10^{14}\,\rm cm^{-3}$ at Earth's surface.
For the parameters applicable to LHe (viz.\ $A=4$, $L_{\rm vap}=21 \,\rm J\,g^{-1}$ , $T_{\rm cry}=4\,\rm K$, ${\bar f}_{\rm KE} = 4/9$)  equation (21) yields fractional vaporization rate of 
$ 3.3 \times 10^{-7} (\sigma_{\rm 300\,K}^{\rm He}/\rm{mb})\,s^{-1} = 3.0\,(\sigma_{\rm 300\,K}^{\rm He} /{\rm mb})\,\% \rm \,\, per\,\,day.$  

Equation (21) relies on the assumption that the mean free path in the cryogen, $\lambda_{\rm cry}$, exceeds the diameter, $D$, of the dewar, so that the entire volume of cryogen is exposed to HIDM at 300~K.  \ma{In the opposite limit, where $\lambda_{\rm cry} \ll D$, HIDM deposit their heat in the outer part of the cryogen and the interior is unheated.  \bzzz{The \be{flux of warm HIDM entering the dewar is $(1/4) n_{\rm DM} (R_\earth) \bar{v}$}, and the resultant energy flux is 
\gr{$(1/4) n_{\rm DM} (R_\earth) \langle \onehalf m_{\rm DM} v^2 v  \rangle_{\rm MB}$}.  The average \gr{fractional energy transfer} is at 
least ${\bar f}_{\rm KE}$ per entering particle, since in this limit they scatter at least once before leaving.}  In this ``opaque dewar" limit, we may set a lower limit on the heating rate per cryogen atom:
$$H_{\rm \be{od}} > {k(T_{\rm DM} - T_{\rm cry})
\,n_{\rm DM} (R_\earth)\,{\bar v}\,\Sigma \over \gr{2}N_{\rm A}}\be{{\bar f}_{\rm KE}}, \eqno(22)$$
where $\Sigma$ is the surface area of the cryogenic volume, and $N_{\rm A}$ is the number of atoms in the dewar.  The latter is given by 
$N_{\rm A} = \rho_{\rm cry} V_{\rm D} / m_{\rm A},$ 
where $V_{\rm D}$ is the volume of the cryogen and $\rho_{\rm cry}$ is its density.  Thus, 
$$H_{\rm \be{od}} > \gr{\frac{1}{2}}\,k(T_{\rm DM} - T_{\rm cry})
\,n_{\rm DM} (R_\earth)\,{\bar v}\, {m_{\rm A} \over \rho_{\rm cry}} {\Sigma \over V_{\rm D}} \be{{\bar f}_{\rm KE}}.  \eqno(23)$$
For a spherical volume, which minimizes the surface area to volume ratio, $\Sigma/V_{\rm D} = 6/D = (36 \pi/ V_{\rm D})^{1/3}.$} 

\ma{For small $\sigma_{\rm 300\,K}^{\rm A}$, the dewar is transparent and $H$ is given by $H_{\rm \be{td}}$, which increases linearly with $\sigma_{\rm 300\,K}^{\rm A}$.  But, once the dewar becomes opaque, $H$ stops increasing linearly \bzzz{with $\sigma_{\rm 300\,K}^{\rm A}$} and approaches $H_{\rm \be{od}}$ asymptotically.  The transparent dewar limit applies for 
$H_{\rm \be{td}} < H_{\rm \be{od}}$.  Given the lower limit on $H_{\rm \be{od}}$ obtained using (\gr{23}) for a spherical cryogenic volume, and comparing it with equation (21), we find that $H_{\rm \be{td}} < H_{\rm \be{od}}$ whenever $\sigma_{\rm 300\,K}^{\rm A} \le 3 m_{\rm A} / (\bu{2} D \rho_{\rm cry})$ or equivalently $\lambda_{\rm cry} < 3D/\bu{2}$.
We may obtain a conservative estimate of $H$ by adopting $H_{\rm \be{td}}$ for $\sigma_{\rm 300\,K}^{\rm A} < 3 m_{\rm A} / (\bu{2} D \rho_{\rm cry})$ and assuming $H$ to be constant for larger $\sigma_{\rm 300\,K}^{\rm A}.$  The corresponding fractional vaporization rate is then obtained by dividing this estimate of $H$ by $L_{\rm vap} m_{\rm A}.$}

\begin{figure}
\includegraphics[width=14 cm]{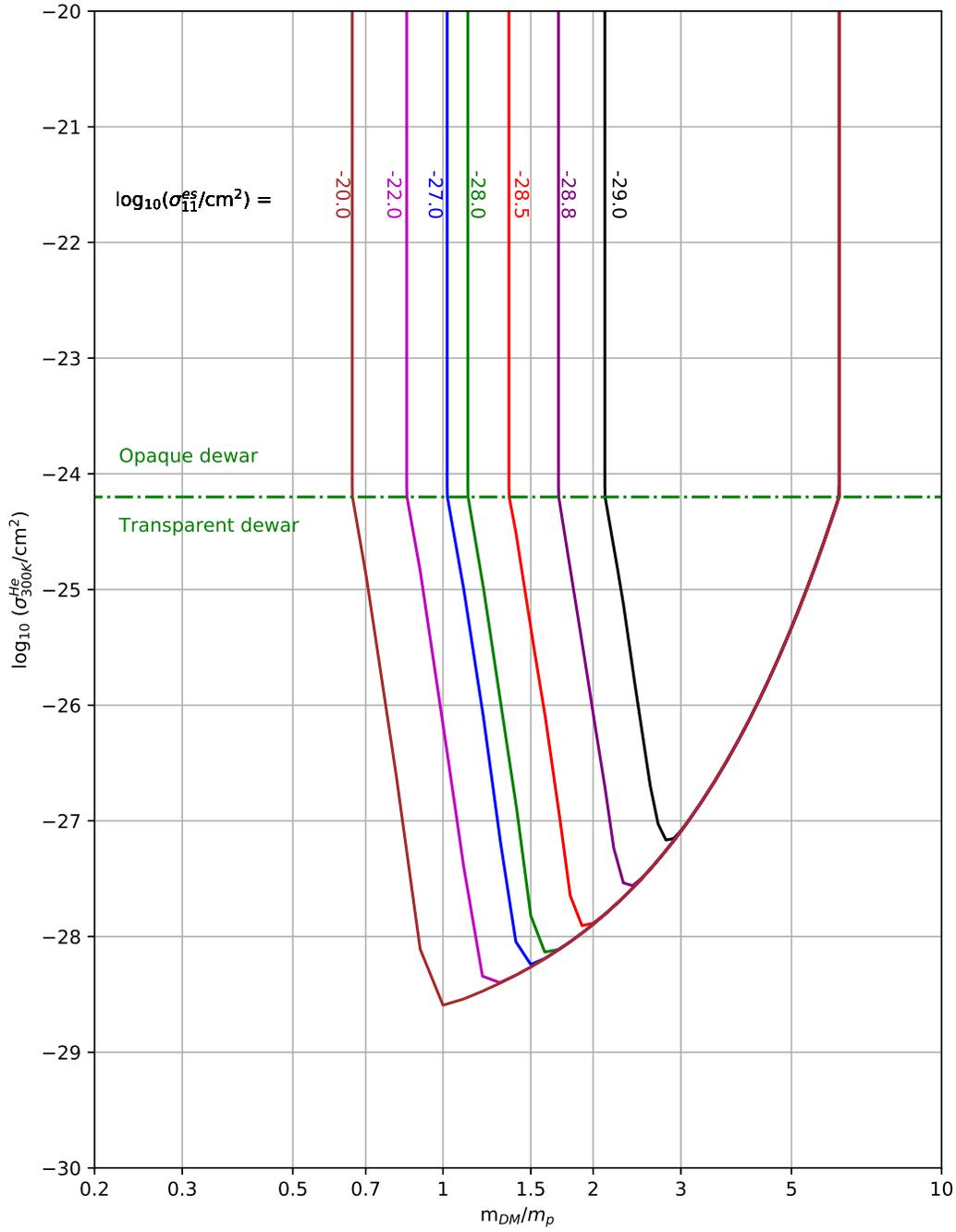}
\caption{Upper limits on $\sigma_{\rm 300\, \rm K}^{\rm He}$ implied by a LHe vaporization rate of $\le 0.5\%$ per day, for LHe exposed to 300~K thermal HIDM.  Results are shown for several values of $\sigma_{11}^{\rm es}$.  The curves are labeled with log$_{10}(\sigma_{11}^{\rm es})$.}
\end{figure}

The most stringent upper limits on $\sigma_{\rm 300\,K}^{\rm A}$ can be obtained 
in the case of liquid He.  Well-designed LHe dewars can achieve fractional boil-off rates as small as 0.5$\%$ per day, when full, without the use of active cooling.\footnote{Specifications sheet for the Cryofab CMSH 1000 LHe Container, downloaded from https://www.cryofab.com/products/ on 2018 March 5}
In Figure 10, we show corresponding upper limits that are implied for $\sigma_{\rm 300\,K}^{\rm He}$.  As was the case for $\sigma_{\rm 6.5\,TeV}^{\rm p, inel}$ (see Section 3.1 above), the limits $\sigma_{\rm 300\,K}^{\rm He}$  depend on what is assumed for $\sigma_{11}^{\rm es}$; once again, we have plotted results obtained for several assumed values of $\sigma_{11}^{\rm es}$.  The results shown here are for a 1000 liter dewar containing LHe with a density of 0.125$\,\rm g \, cm^{-3}.$

Figure 10 also assumes that HIDM particles can enter the LHe within the dewar without scattering first within the multilayer insulation (MLI) that is typically used to mimimize the radiative heat load.  
A very well-insulated dewar might employ \be{several tens} of 
layers of $\bx{\sim 10}\,\micron$ thick aluminized mylar, ($\rm C_{10}H_8O_4)_n$, corresponding to a total mylar thickness of \bx{several hundred $\mu$m} (with a negligible additional thickness of aluminum).  Thus, our analysis is only valid if the HIDM mean-free path in mylar, $\lambda_{\rm mylar}$, exceeds \bx{that thickness}, or equivalently if the mean cross-section per atom in mylar is less than $\sim {\be{\rm few}} \times 10^{-22}$~cm$^2$.  \be{Because orbiting spacecraft, including HST, are often covered with similar multilayer insulation for thermal control, a cross-section any larger than this value would mean that the opaque spacecraft limit applies (Section 3.2); this is turn would rule out a large portion of the available parameter space \bx{anyway} (dashed blue contour in Figure 8)}.
One caveat applies to this argument: the HIDM striking a spacecraft have a relative velocity of $\sim 8 \,\rm km \, s^{-1}$, while those incident on the dewar insulation have a velocity of $2.51 \,(m_{\rm DM}/m_{\rm p})^{-1/2}\rm \, km \, s^{-1}$.  Thus, if the cross-section were \bzzz{a sufficiently} strongly decreasing function of collision velocity, this argument \be{would not necessarily apply.}

\begin{table}
\caption{\ma{Parameters for cryogenic liquids}}
\vskip 0.1 true in
\begin{tabular}{lcccccc}
\hline
Cryogen & $T_{\rm cry}$ & $\rho_{\rm cry}$ & $L_{\rm vap}$ & $V$ & $V_{\rm D}$ & Reference \\
        & K             & g~cm$^{-3}$ 	   & J g$^{-1}$    & \% per day & L & \\
\hline 
He		& 4 & 0.125 & 21 & 0.5 & 1000 & (1) \\	
H$_2$	& 20 & 0.071 & 446 & 0.26 & 500 & (2) \\
N$_2$	& 77 & 0.81 & 199 & 1.3 & 300 & (3) \\
O$_2$	& 90 & 1.14 & 212 & 0.8 & 300 & (3) \\
Ar		& 97 & 1.40 & 162 & \be{0.8} & 300 & (3) \\	
\hline 
\multicolumn{5}{l}{References: (1) footnote 4; (2) Birmingham et al.\ 1957;} \\
\multicolumn{5}{l}{ (3) same as (1), but for the CLD 300 tank}\\
\end{tabular}
\end{table}

Additional limits may be obtained for other nuclei by considering the vaporization of other cryogens.  Figure 11 shows analogous limits for H, N, O, and Ar.  These were obtained from limits on the boil-off rates for storage dewars for liquid H$_2$, N$_2$, O$_2$ and Ar.  The relevant parameters for these cryogens are tabulated in Table 1. For nuclei where no liquid cryogen is available, constraints on $\sigma_{\rm 300\,K}^{\rm A}$ could be obtained by experiments in which solid materials containing the nucleus in question are placed in a storage dewar and immersed in a liquid cryogen.  Any heat deposited by HIDM within the solid material \be{would} be transferred to the cryogen; this effect could be detected by means of a differential measurement in which the boil-off rates were compared for cases with and without an immersed sample of the material to be tested.

\begin{figure}
\includegraphics[width=14 cm]{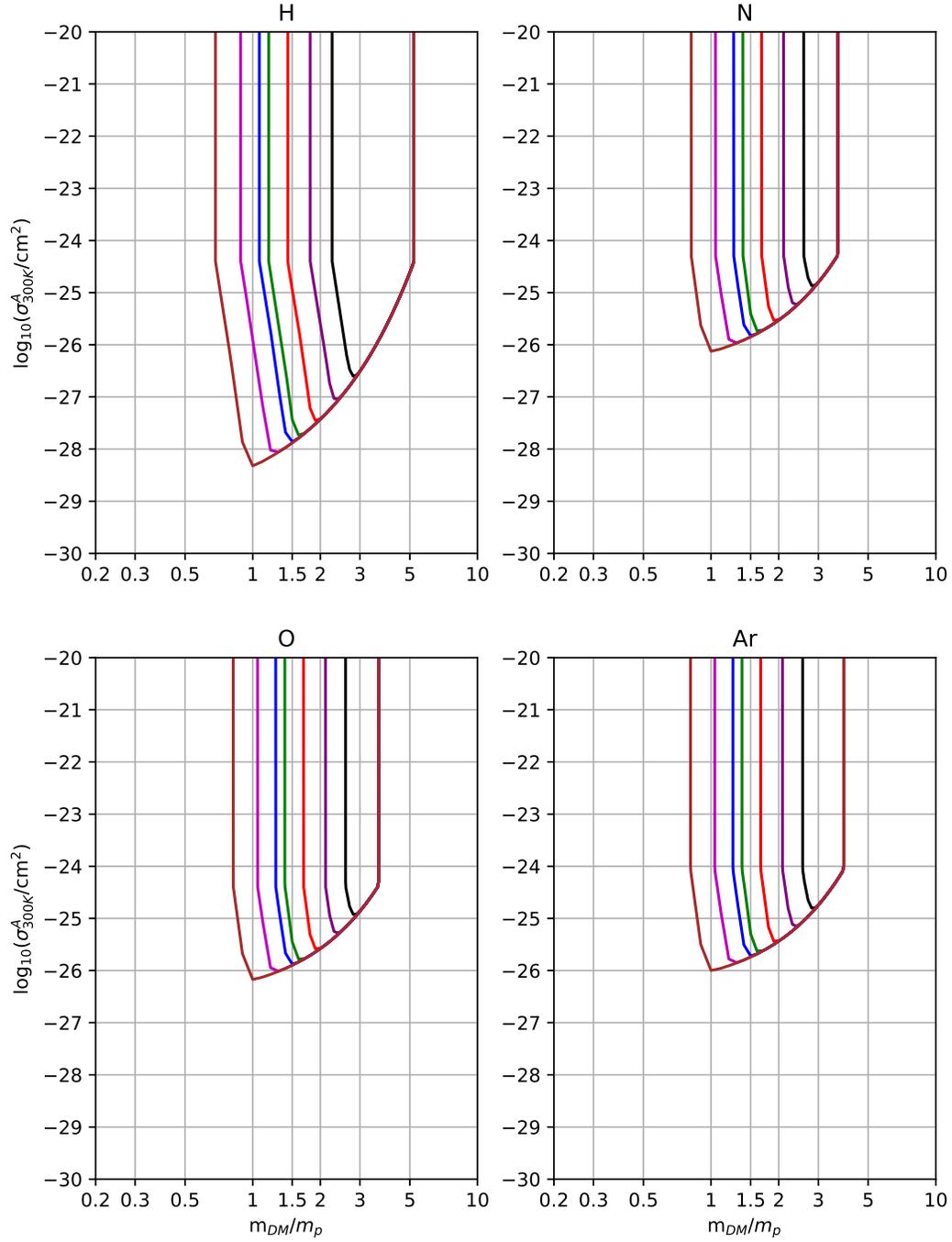}
\caption{Same as Figure 10, but for the other cryogens and boil-off rates listed in Table 1.}
\end{figure}

The limits presented in Figures 10 and 11 are very conservative, in that they assume the observed boil-off rates to be entirely attributable to scattering of thermal HIDM.  Improved limits could be obtained by measurements of how the mass-loss rate of cryogen, $\dot{M}_{\rm cry}$, depends on the mass of cryogen, $M_{\rm cry}$, present within the dewar.  The expected mass-loss rate of cryogen 
is $$\dot{M}_{\rm cry} = B_{\rm DM} M_{\rm cry} + Q / L_{\rm vap}, \eqno(24)$$
where $Q$ is the heat leak in the dewar and $B_{\rm DM}$ is the fractional boil-off rate due to HIDM heating.  \bzzz{Figures 10 and 11 are conservatively based upon} $B_{\rm DM} \le \dot{M}_{\rm cry}/M_0,$ where $\dot{M}_{\rm cry}$ is determined for full dewar containing a mass $M_0$ of cryogen.  To the extent that the inner vessel within the dewar is isothermal and at the boiling point of the cryogen, the heat leak, $Q$, may be expected to be independent of $M_{\rm cry}$, provided any cryogen remains within the dewar.  Thus, a limit can be placed on $B_{\rm DM}$ by comparing the mass-loss rates for a full dewar containing a cryogen mass $M_0$ and a nearly-empty dewar.  If the difference in the two mass loss-rates is less than $\delta \dot{M}_{\rm cry}$, then the limit on $B_{\rm DM}$ becomes $B_{\rm DM}  < \delta \dot{M}_{\rm cry}/M_0$ instead of $B_{\rm DM} \le \dot{M}_{\rm cry}/M_0$.  If the mass-loss rates for full and nearly-empty dewars could be demonstrated to differ by less than 1$\%$, for example, the limit would improve by a factor of 100.

\subsection{The thermal conductivity within the Earth's crust}

The effects of HIDM on cryogenic experiments are one manifestation of heat transport. A related consequence of captured HIDM would be to enhance the thermal conductivity in the Earth's crust.  \be{Similar effects have been considered for the case of weakly-interacting massive particles within the Sun (Spergel \& Press 1985; \bz{Faulkner \& Gilliland 1985)}.}  
\bzzz{In the regime of present interest, the HIDM constitute a monoatomic gas that is ``Lorentzian" in the terminology of CC70 (Section 10.5), meaning that $m_{\rm DM}$ is small compared to the mean atomic mass of crustal materials, $21.5\,m_{\rm p}$, and $n_{\rm DM}$ is much smaller than the number density $n_{\rm cr}$, of atoms in the crust.}  For a mean scattering cross-section, $\sigma^{\rm cr}_0$, that is independent of the
collision velocity, the additional thermal conductivity associated with such a gas was first computed by Lorentz (1904) and may be written\footnote{Here, in using a thermal conductivity to describe heat transport in the crust, we implicitly assume that the \bzzz{thermalization length} $\lambda_*$
is much smaller than the length scale on which the temperature varies \bzzz{(see discussion in Section 2.3 above)}.}
$$k_{\rm DM}={2 n_{\rm DM} k {\bar v} \lambda \over 3} = 
{2 n_{\rm DM} k {\bar v} \over 3  n_{\rm cr} \sigma^{\rm cr}_0}.\eqno(25)$$ 
An important feature of this expression is that $k_{\rm DM}$ is a {\it decreasing} function of the cross-section; for a given value of $n_{\rm DM} (R_\earth)$, an upper limit on the allowable value of $k_{\rm DM}$ therefore sets a {\it lower} limit on $\sigma^{\rm cr}_0.$

\begin{figure}
\includegraphics[width=13 cm]{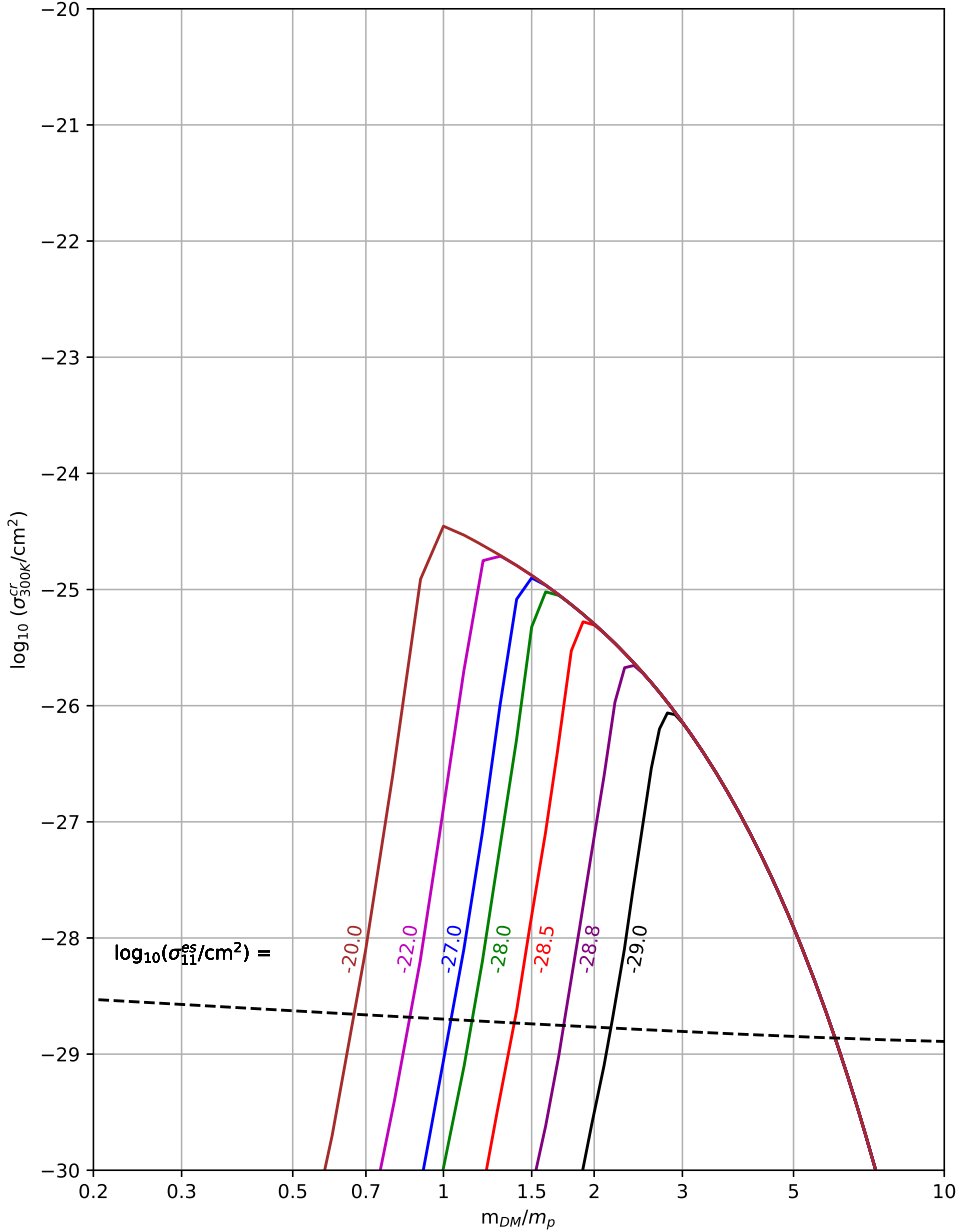}
\caption{Lower limits on $\sigma^{\gr{\prime} \rm cr}_{\rm 300\, K}$, obtained from the requirement $k_{\rm DM} \le \bu{4.3 \times 10^{5}}\, \rm erg \, s^{-1} \,cm^{-1} \, K^{-1}$. 
Results are shown for several values of $\sigma_{11}^{\rm es}$.  The curves are labeled with log$_{10}(\sigma_{11}^{\rm es}/{\rm cm}^2).$  \be{The allowed region is {\it above} the curves.}   \byy{The criterion for the validity of the heat conduction equation, $\lambda_* < (d\ln T/dz)^{-1} \sim 10$~km, is met above the dashed line in Figure 12.  A sufficient condition for the validity of the treatment in Appendix C, $\lambda_* < 100$~m, is met for cross-sections more than two orders of magnitude above the dashed line.}}
\end{figure}

\bzz{CC70 generalized the treatment of the conductivity of a Lorentzian gas to the case where the scattering force obeys an inverse power-law dependence, $F \propto r^{-q}$, and the 
momentum-transfer cross-section therefore has a power-law dependence on velocity.  As in Section 3.3 above, we write this cross-section in the form $\sigma_{\rm v}^{\rm cr} = \sigma_{\rm 0}^{\rm cr} (v/v_0)^{-j}$.  Upon comparison with CC70 equation (10.31), in which the cross-section is denoted $\phi^{(1)}_{12}$, we find that the power-law indices $j$ and $q$ are related according to $j = 4/(q-1)$.   For this case, the thermal conductivity is given by CC70 equation (10.51,5), which we may rewrite in the form  
$$k_{\rm DM}= {2 n_{\rm DM} k {\bar v} \over 3 n_{\rm cr} \sigma^{\prime \rm cr}_{\rm T} },\eqno(26)$$
where 
$$\sigma_{\rm T}^{\prime \rm cr} = {2 \over \Gamma(3+\onehalf j)} (2 k T/[m v_0^2])^{-j/2} \sigma^{\rm cr}_{0}\eqno(27)$$
is an effective cross-section that replaces 
$\sigma^{\rm cr}_0$ in equation (25).  This cross-section is analogous to the Rosseland mean cross-section in the theory of radiative diffusion (Rosseland 1925) -- being an appropriately-weighted harmonic mean  -- and reduces to $\sigma^{\rm cr}_{0}$ in the velocity-independent case ($j=0$). 
Putting in numerical values, we then obtain 
$$ k_{\rm DM} = \bzz{3.1} \times 10^4 {\rm \,erg \, cm^{-1} \, K^{-1}}\, 
\biggl({T_{\rm DM}  \over 300\,\rm{K}} \biggr)^{\gr{1/2}} 
\biggl({m_{\rm DM}  \over m_{\rm p}} \biggr)^{-1/2}
\biggl({n_{\rm DM} (R_\earth) \over 10^{14}\,\rm cm^{-3}} \biggr)
\biggl({\sigma^{\rm \prime cr}_{\rm T} \over 10^{-24} \, \rm cm^{2}} \biggr)^{-1}.
 \eqno(28)$$}
Here, we adopted a mean density of 2.7$\,\rm g\,cm^{-3}$ for the crust (Dziewonski \& Anderson 1981), and a mean atomic mass of $21.5\,m_{\rm p}$, \be{yielding} an atomic number density $n_{\rm cr} = 7.5 \times 10^{22} \rm cm^{-3}$.  The mean free path, which is determined by $n_{\rm  cr}$, not $n_{\rm DM}$, is therefore $\lambda=13.3\,(\sigma_{\rm T}^{\prime \rm cr}/10^{-24} \, \rm cm^{2})^{-1}\,\rm cm$. 

Comparing this to the analogous expression obtained for $\sigma_{\rm T}^{\rm A}$ in Section 3.1 above, we obtain the following ratio of the cross-section that determines the conductivity to the mean cross-section that is relevant to the \be{vaporization} of cryogens: $\sigma_{\rm T}^{\prime \rm A}/\sigma_{\rm T}^{\rm A} = 4 / [\Gamma(3+\onehalf j) \Gamma(3-\onehalf j)].$  For $j=0$ (i.e.\ with no velocity-dependence), $\sigma_{\rm T}^{\prime \rm A}/\sigma_{\rm T}^{\rm A} = 1$ as required.  For $j=4$ (velocity-dependence for Rutherford scattering), $\sigma_{\rm T}^{\prime \rm A}/\sigma_{\rm T}^{\rm A} = 1/6$.

Heat flow through the Earth's crust has been probed extensively through measurements of the varying temperature gradients in boreholes.  \byy{Such measurements have led to an estimate of $47\pm 2$~TW for the total power transported through the Earth's crust (Davies \& Davies 2010).  A value of $k_{\rm DM}$ any larger than ${3.7 \times 10^5}\, {\rm erg \, cm^{-1} \, K^{-1}}$ would more than double current estimates for the global heat flow from the Earth, and would require significant modifications to the standard understanding of the Earth's internal heat budget (e.g.\ Lay et al.\ 2008).}

Moreover, in cases where a borehole probes a stratified series of rock formations, changes in the temperature gradient can be sometimes be discerned at the boundaries of the different rock types (e.g. Harris \& Chapman 1995; hereafter HC95).  The observed behavior is found to be in quantitative agreement with the ratios of conductivities that are measured in the laboratory for the different rocky materials involved, 
placing limits on any anomalous conductivity associated with HIDM. 
A more detailed discussion of such measurements appears 
in Appendix C, where we derive an upper limit of $4.3 \times 10^5\, {\rm erg \, cm^{-1} \, K^{-1}}$ on $k_{\rm DM}$ \byy{for cross-sections such that the thermalization length, $\lambda_*$ (see section 2.3 above), is less than the characteristic length scale $\sim 100$~m probed in HC95.}  
Figure 12 shows the {\it lower} limits on $\bz{\sigma^{\prime \rm cr}_{\rm 300\, K}}$ implied by this constraint, \byy{which are very similar to the limits imposed the global heat flow constraint, $k_{\rm DM} < {3.7 \times 10^5}\, {\rm erg \, cm^{-1} \, K^{-1}}$, under the assumption that heat transport by HIDM does not increase the latter by more than a factor 2.}  As in Figures 7, 9, and 10, the results depend on what is assumed for $\sigma_{11}^{\rm es}$; once again, we have plotted results obtained for several values of $\sigma_{11}^{\rm es}$.   \byy{The criterion for the validity of the heat conduction equation, $\lambda_* < (d\ln T/dz)^{-1} \sim 10$~km, is met above the dashed line in Figure 12.  A sufficient condition for the validity of the treatment in Appendix C, $\lambda_* < 100$~m, is met for cross-sections more than two orders of magnitude above the dashed line.}


\section{Summary and conclusions}

The various considerations described in Section 3 provide constraints on the properties of the HIDM.   \bzzz{These constraints
are based upon a model for the number of HIDM particles that have been captured and retained over the lifetime of the Earth and for 
their resultant density distribution.   
Our treatment of the latter makes use of a differential equation for the partial pressure of HIDM particles
(eqn.\ 8) that may be derived from the collisional Boltzmann equation (CC70) and was adopted by Gilliland et al.\ (1986) 
in their study of WIMP dark matter in the Sun.  As we have discussed in Appendix A, a modification to that equation proposed
subsequently by GR90 violates considerations of hydrostatic equilibrium and presents other pathological behaviors.}

\bx{\bzzz{We note that} our discussion of the density of HIDM particles within the Earth, and of the resultant constraints presented in Section 3, assumes that HIDM \bzzz{does not annihilate with nucleons in the Earth (limited by Farrar and Zaharijas 2006 and Mack et al.\ 2007) and is not destroyed  by self-annihilation.}  The latter assumption is valid provided the self-annihilation 
cross-section is smaller than 
$$\sigma^{\rm sa}_{\rm crit} = {N_{\rm C} \over \int_0^{R_\earth} 4 \pi r^2 n_{\rm DM}^2 {\bar v} t_\earth dr}, \eqno(29)$$ where $N_{\rm C}$ is the number of captured particles.  The critical value of the self-annihilation cross-section, $\sigma^{\rm sa}_{\rm crit}$, above which self-annihilation reduces $n_{\rm DM}$ and weakens the limits set in Section 3, ranges from $1.0 \times 10^{-36}\,\rm cm^{2}$ to $1.0 \times 10^{-34}\,\rm cm^{2}$ as $m_{\rm DM}$ ranges from 1 to 5 $m_{\rm p}$. }

Because the constraints we derive involve cross-sections for multiple baryonic nuclei at a variety of collision energies , and because the cross-sections may show strong non-monotonic variations with nucleon number, $A$ (\byy{Farrar \& Xu 2018)}, the constrained 
parameter space has a high dimensionality.  Several different cross-sections appear in the various 
constraints obtained in Section 3: 
$\sigma_{\rm 11}^{\rm cr}$, 
$\sigma_{\rm 11}^{\rm atm}$, 
$\sigma_{\rm 6.5 \,TeV}^{\rm p}$, 
$\sigma_{\rm 8}^{\rm HST}$, 
$\sigma_{300\,\rm K}^{\rm He}$, 
$\sigma_{300\,\rm K}^{\rm H}$, 
$\sigma_{300\,\rm K}^{\rm N}$, 
$\sigma_{300\,\rm K}^{\rm O}$, 
$\sigma_{300\,\rm K}^{\rm N}$,
$\sigma_{300\,\rm K}^{\rm cr}$, \be{$\sigma_{300\,\rm K}^{\prime \rm cr}$},
$\sigma_{1}^{\rm p}$,
$\sigma_{\rm 11}^{\rm p}$,
$\sigma_{\rm 11}^{\rm He}$.
Any proposal for the HIDM that makes a specific prediction for the $A$-dependence and velocity dependence for the scattering cross-section with baryons can be evaluated with respect to the multiple constraints given here.  

Figures 7 and 9 -- 12 provide the full set of constraints that we have obtained, \bzzz{and} the key results \ma{are summarized in Table 2.   These results rely on the assumption that our estimates of the HIDM accumulation and evaporation rates apply throughout Earth's history.  However, they are conservative in the sense that we have derived them under the assumption that HIDM are entirely responsible for limiting the LHC beam lifetime, for the vaporization of liquid cryogens, and for the decay of spacecraft orbits.  Stronger limits could be derived, for example, by modeling LHC beam losses due to conventional effects such as beam particle scattering off residual atmospheric gases in the beam pipe and interactions within the beam or with components of the beam-line.   Or the boil-off of liquid cryogens could be monitored as described in the last paragraph of Section 3.3 }   \bzzzz{All the upper limits presented in Table 2 are inversely proportional to the value adopted for $\rho_{\rm DM}$, the mass density of the dark matter in the Galactic plane, while the lower limits obtained for $\sigma^{\prime cr}_{\rm 300}$~K are proportional to $\rho_{\rm DM}$.  Thus, if $\rho_{\rm DM}$ were larger than the conservative value of $0.3 \,(\rm{GeV}/c^2) \,{\rm cm}^{-3}$ that we adopted, all the constraints we obtained would be strengthened.}

\begin{deluxetable}{llccccccc}
\tabletypesize{\small}
\tablecaption{Constraints$^a$ applying for any $\sigma_{11}^{\rm es}$ in the $10^{-28} - 10^{-20}\, \rm cm^{2}$ range}
\startdata
\hline
Constraint &for $m_{\rm DM}$ =  & 1.0$\, m_{\rm p}$ & 1.2$\, m_{\rm p}$ & 1.5$\, m_{\rm p}$ & 2.0$\, m_{\rm p}$ & 2.5$\, m_{\rm p}$ & 3.0$\, m_{\rm p}$ & 5.0$\, m_{\rm p}$ \\
\hline
LHC beam & $\sigma_{6.5 \rm \, TeV}^{\rm p,inel}$~$<$ &
 69~mb & 1.0~mb	& $1.25\,\mu$b & $1.0\,\mu$b & $2.4\,\mu$b & $6.0\,\mu$b & 0.28~mb \\
 \\
HST orbit & $\sigma_{8}^{\rm HST}$~$<$ & 
 650~mb & 13~mb & $29\,\mu$b & $64\,\mu$b & 0.44~mb & 3.3~mb & $-$ \\
 \\
LHe boil-off & $\sigma_{\rm 300 K}^{\rm He}$~$<$  & 
 $-$ & 120~mb & 0.15~mb & 0.13~mb & 0.31~mb & 0.81~mb & 47~mb \\
LH$_2$ boil-off & $\sigma_{\rm 300 K}^{\rm H}$~$<$  & 
 $-$ & 250~mb & 0.37~mb & 0.37~mb & 1.1~mb & 3.1~mb & 240~mb \\
LN$_2$ boil-off & $\sigma_{\rm 300 K}^{\rm N}$~$<$  & 
 $-$ &  $-$   & 39~mb   & 30~mb & 66~mb & 160~mb &  $-$   \\
LO$_2$ boil-off & $\sigma_{\rm 300 K}^{\rm O}$~$<$  & 
 $-$ &  $-$   & 35~mb   & 26~mb & 58~mb & 140~mb &  $-$  \\
LAr boil-off & $\sigma_{\rm 300 K}^{\rm Ar}$~$<$  & 
 $-$ &  $-$   & 51~mb   & 37~mb & 79~mb & 180~mb &  $-$   \\
 \\ 
Conductivity$^b$ & \bzz{$\sigma_{\rm 300 K}^{\rm \prime cr}$~$>$}  
&  $-$ & \bzzz{$68\,\mu$b} & \bzzz{48~mb} & \bzzz{51~mb} & \bzzz{19~mb} & \bzzz{7.2~mb} & \bzzz{$124\,\mu$b }\\
 \hline
\multicolumn{9}{l}{$^a$ Limits for $m_{\rm DM} = 1.0\,m_{\rm p}$ apply only if $\sigma_{11}^{\rm He}$ and $\sigma_{11}^{\rm H}$ are both $\simlt 1$~b, such that thermospheric} \\
\multicolumn{9}{l}{loss is negligible.  Constraints for $m_{\rm DM} \ge 1.2\, m_{\rm p}$ apply for all $\sigma_{11}^{\rm He}$ and $\sigma_{11}^{\rm H} \le 10^4\,\rm b$. We assume} \\ 
\multicolumn{9}{l}{that thermospheric temperatures in the recent past are typical of
those throughout Earth's history.}  \\
\multicolumn{9}{l}{\bzzzz{$^b$  $\sigma_{\rm 300 K}^{\rm \prime cr}$ is the harmonic average cross-section defined by equation (27)}} \\
\enddata 
\end{deluxetable}

As an illustrative example \bzzz{of the application of our constraints}, we consider a case where the cross-section has a $v^{-4}$ dependence on collision velocity, $v$.  Such a velocity dependence was considered by Mu{\~n}oz et al.\ (2015) in their analysis of the cooling of hydrogen atoms at high redshift, an analysis that was recently invoked (Barkana 2018) as the explanation for an anomalously deep 21 cm absorption feature reported at \bzzz{$z \sim 17$} (Bowman et al.\ 2018).\footnote{{\bx{Although subsequent analyses have cast doubt on both the observations (e.g.\ Hills et al.\ 2018) and the interpretation offered to explain them (e.g.\ \bz{Mu{\~n}oz \& Loeb 2018}; Berlin et al.\ 2018; Barkana et al.\ 2018), this example remains a useful illustration of how the constraints we have presented here may be applied in the context of a specific DM candidate.}} } 
In this example, we assume that the scattering cross-sections \be{have a Rutherford-like behavior, i.e.\ are} proportional to $(A / [\mu_{\rm A} v^2])^2,$ as discussed in Section 2.1.  \be{Then} the scattering cross-sections \be{only depend on a single parameter, $\sigma_1^{\rm p}$}, via  $\sigma_{\rm v}^{\rm A} = (A \mu_{\rm p}/\mu_{\rm A})^2 (v/\rm km\,\,s^{-1})^{-4}\sigma_1^{\rm p}$, where $\sigma_1^{\rm p}$ is the cross-section for collisions with protons at $v = 1 \rm \, km\,s^{-1}$.  For $m_{\rm DM}$ in the range $0.5 - 5\, m_{\rm p}$, $\sigma_{\rm v}^{\rm cr}/\sigma_{\rm v}^{\rm p}$ varies from $237 - 20.8$ and $\sigma_{\rm v}^{\rm atm}/\sigma_{\rm v}^{\rm p}$ varies from $102 - 10.7$ \bzzz{due to the variation of $\mu_{\rm p}/\mu_{\rm A}$ with $m_{\rm DM}$.}
To yield a significant cooling effect on hydrogen atoms at high redshift, $\sigma_1^{\rm p}$ values of order $10^{-19} - 10^{-18} \, \rm cm^2$ are required (Mu{\~n}oz et al.\ 2015; Barkana 2018).  In this example, we therefore focus on $\sigma_1^{\rm p}$ in the range $10^{-20} - 10^{-16} \, \rm cm^2$.  \be{For this range of cross-sections}, incident \be{DM} particles at $v = 200 \rm \, km\,s^{-1}$ are first scattered in the crust at depths $\gr \le 1$~km and may therefore be captured.  Furthermore, the LSS (for \be{DM} particles with $v \sim v_{\rm es} = 11.2 \rm \,km\,s^{-1}$) is below an altitude of 100~km and HST is opaque to HIDM at $v_{\rm orb} \sim 8 \, \rm km\,s^{-1}$.  In this region of parameter space, equation (19) may be used to estimate the orbital decay rate, $-dR_{\rm orb}/dt$, of HST.  In Figure 13, we plot $-dR_{\rm orb}/dt$ for HST as a function of $m_{\rm DM}$, with the horizontal green line indicating the value,  $-dR_{\rm orb}/dt=0.8 \, \rm km \, yr^{-1}$, experienced by HST between servicing missions SM1 and SM2.   The results plotted here include the effects of thermospheric loss.  Because the $A$- and velocity-dependences are specified in this example, the relevant cross-sections for thermospheric escape ($\sigma_{11}^{\rm He}$ and $\sigma_{11}^{\rm H}$) have a fixed relationship to $\sigma_1^{\rm p}$, as do all the cross-sections of relevance in our model.  For $\sigma_{1}^{\rm p}$ in the range proposed by Barkana (2018), $8 \times 10^{-20} - 10^{-18} \, \rm cm^2$, masses in the range \be{0.55 - 3.9}~$m_{\rm p}$ are ruled out.\footnote{\be{In his treatment of scattering by primordial material at $z \sim 17$, Barkana (2018) assumed a $\sigma_{\rm v}^{\rm A} \propto A$  dependence that was different from that adopted here (personal communication).  For this A-dependence, the range of excluded masses is very similar: $0.60 - 3.9\,m_{\rm p}$}}. 

\begin{figure}
\includegraphics[width=18 cm]{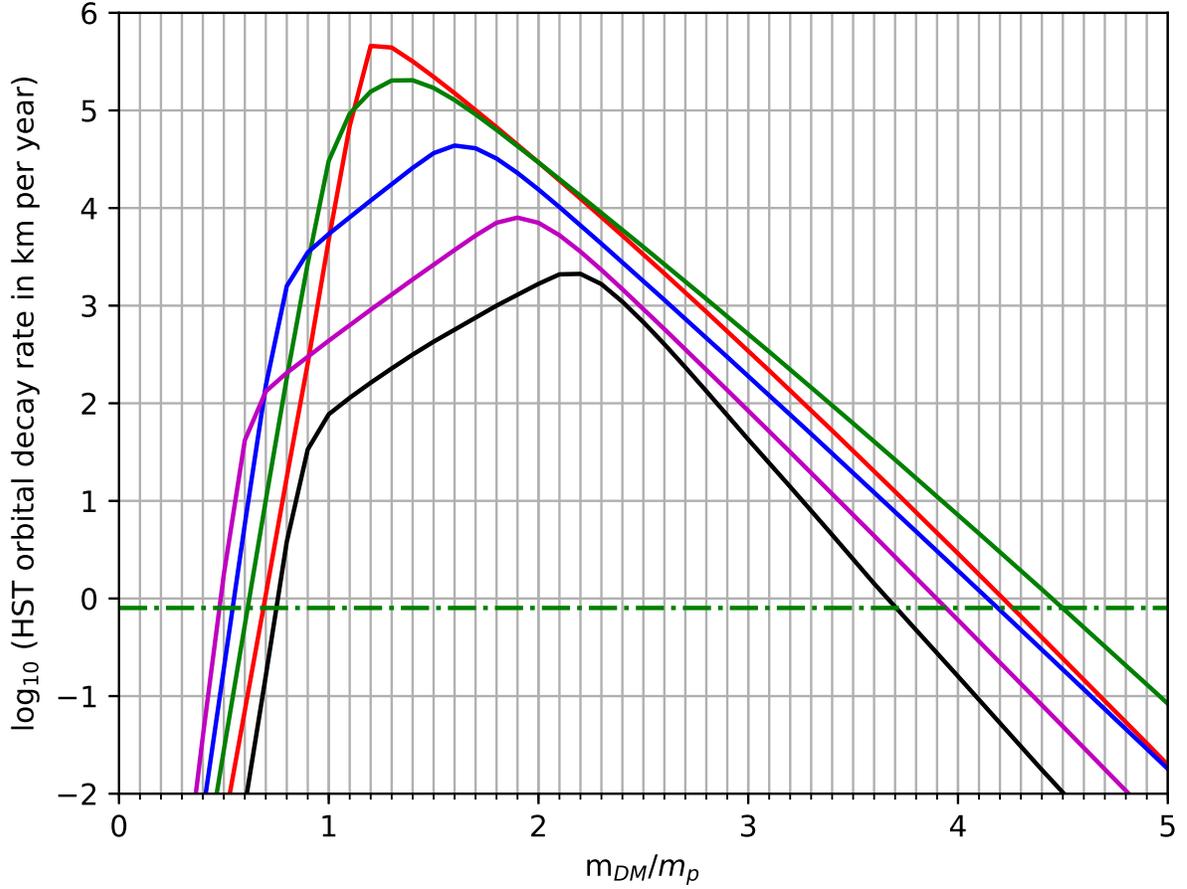}
\caption{Orbital decay rate of HST, $-dR_{\rm orb}/dt$, for a scattering cross-section $\sigma_{\rm v}^{\rm A} = (A \mu_{\rm p}/\mu_{\rm A})^2 (v/\rm km\,\,s^{-1})^{-4}\sigma_1^{\rm p}$. 
The horizontal green line indicates the value experienced by HST between servicing missions SM1 and SM2: $dR_{\rm orb}/dt= - 0.8 \, \rm km \, yr^{-1}$.  \be{Red, green, blue, magenta and black curves apply to $\sigma_1^{\rm p}$ values of $10^{-21}$, $10^{-20}$, $10^{-19}$, $10^{-18}$, and $10^{-17}\,\rm cm^2$ respectively.}}
\end{figure}

\byy{To summarize, we have shown that for dark matter in the $0.60 - 6 \,m_{\rm p}$ mass range, having a momentum-transfer cross-section $\simgt 10^{-29}\rm \, cm^2$ for scattering with material in the crust, the Earth will have a significant atmosphere of dark matter extending throughout the interior and far above the surface.  We determined the density and structure of this DM atmosphere from first principles, finding that the density can exceed $10^{14}\, \rm cm^{-3}$ at the Earth's surface.   Given this high density, we infer upper limits on scattering cross-sections that are generally stronger than those from direct detection experiments, using bounds on the orbital decay of HST and the evaporation of liquid cryogens.  These upper limits are complemented by lower limits from the thermal conductivity of the Earth’s crust, which provide a stringent constraint on models, especially when the Born approximation can be used to relate the cross sections for different nuclei.  Typical DM velocities at the surface in such an atmosphere are $\sim \rm few \, km \,s^{-1}$, providing strong constraints on models with 
a $\sigma_{\rm v}^{\rm A} \propto v^{-4}$ behavior at low velocity.  
Finally, we have obtained the first limits on the high energy HIDM-proton scattering cross-section: $\sigma_{\rm p} ^{\rm inel} \le  9 \times 10^{-31}\,  [10^{14}{\rm cm}^{-3}/n_{\rm DM}(R_\earth)]\, {\rm cm}^2$.}

\begin{acknowledgements}

We all gratefully acknowledge the hospitality of the Center for Computational Astronomy of the Flatiron Institute in New York, where D.A.N.\ and C.F.M.\ were visiting scholars when this study was initiated.  We thank Daniel Brach-Neufeld, Raymond Jeanloz, Bob Johnson, \be{Chris Long}, Michael Manga, Toby Marriage, Lyman Page, \bzzz{David Spergel}, \be{Rai Weiss and Xingchen Xu} for helpful discussions.  The work of C.F.M.\ was supported \bu{in part} by NSF grant AST-1211729, \be{and that of G.R.F. was supported by NSF grant AST-1517319.}

\end{acknowledgements}

\vfill\eject

\appendix

\section{Density distribution of DM}

Based on an analysis of the Boltzmann collision equation for a dilute gas of particles of mass, $m_{\rm X}$, within a ``background" gas of atoms of mass, $m_{\rm n}$,
Gould \& Raffelt (1990; hereafter GR90) proposed a modification to the density distribution adopted by Gilliland et al.\ (1986) for DM in the Sun in the limit of large scattering cross-section.
Whereas Gilliland et al.\ (1986) adopted a particle density profile given by
$${n_{\rm DM}(r) \over  n_{\rm DM}(0)} = \biggl( {T(r) \over T(0)} \biggr)^{-1} \, \exp \biggl( - \int {m_{DM} d\Phi \over kT(r)} \biggr), \eqno(A1)$$
which is the integral form of our equation (8) and is in agreement with that implied by CC70 equation (8.1,7), 
GR90 obtained the expression (their equation 2.30)
$${n_{\rm DM}(r) \over n_{\rm DM}(0)} = \biggl( {T(r) \over T(0)} \biggr)^{3/2} \, \exp \biggl( - \int \biggl[ \alpha {d{\rm ln T} \over dr} + 
{m_{DM} g \over kT(r)}\biggr] dr \biggr). \eqno(A2)$$
Here, the dimensionless quantity $\alpha$, which was computed numerically and tabulated by GR90, was found to be a monotonically increasing function of $m_{\rm X}/m_{\rm n}$, having values of 2, 2.32, and 2.5 respectively for $m_{\rm X}/m_{\rm n}=$ 0, 1, and $\infty$.  This modification amounts to the addition of a term $({5 \over 2} -\alpha)d\ln T(r)/dr$ to the right-hand-side of our equation (8), yielding (GR90; equation 2.29)
$${d{\rm ln}p_{\rm DM} \over dr} = - {m_{\rm DM} g\be{(r)} \over kT\bx{(r)}} + 
(5/2 - \alpha){d{\rm ln}T(r) \over dr}. \eqno(A3)$$
 
In the limit of large $m_{\rm X}/m_{\rm n}$, the additional term $({5 \over 2} -\alpha)d\ln T(r)/dr$ is zero and the DM density distribution is identical to that adopted by Gilliland et al.\ (1986).  This is also apparent from the form of equation (A2) given by GR90 for the case of constant $\alpha$
(their equation 2.31)
$${n_{\rm DM}(r) \over  n_{\rm DM}(0)} = \biggl( {T(r) \over T(0)} \biggr)^{(3/2-\alpha)} \, \exp \biggl( - \int {m_{DM} g \over kT(r)} dr \biggr). \eqno(A4)$$
However, outside the limit of large $m_{\rm X}/m_{\rm n}$, the partial pressure gradient $d{\rm ln}p_{\rm DM}/dr$ given by GR90's treatment is different from that given by equation (8); as we discuss below, this would appear to present an inconsistency with the hydrostatic equilibrium equation, $dp_{\rm tot}/dr=- g \rho_{\rm tot},$ where $p_{\rm tot}$ and $\rho_{\rm tot}$ are the total pressure and density.

As a thought experiment, let us consider a single component gas in hydrostatic equilibrium, and label one in every million particles and consider them to comprise
the dilute gas under consideration.   
The partial pressure, $p_{\rm X}$, and density, $\rho_{\rm X}$, 
for that dilute gas is everywhere a factor of one million \bzzz{times smaller than} the total pressure and density, and must therefore also obey the hydrostatic equilibrium equation, i.e. $dp_{\rm X}/dr=- g \rho_{\rm X}.$  But this equation is exactly equation (8), without the additional term proposed by GR90, and thus the required value of $\alpha$ is 5/2, not the 2.32 given by GR90 for the case $m_{\rm X}/m_{\rm n}=1$

Similar inconsistencies arise for $m_{\rm X}/m_{\rm n} \ne 1.$  To demonstrate this point, let us return to the binary gas mixture consisting of a dilute gas of particles of mass $m_{\rm X}$ within a ``background" gas of atoms of mass, $m_{\rm n}$, and now consider the case where there is a temperature gradient but no gravitational field.   In equilibrium, the total gas pressure must be constant and thus the total particle number density, $n_{\rm X} + n_{\rm n}$, must be inversely proportional to temperature.  However, GR90 equation (2.29) tells us that the density of the dilute gas is $n_{\rm X} \propto T^{(\alpha - 3/2)}$, and the concentration of the dilute gas, $n_{\rm X}/(n_{\rm X} + n_{\rm n})$, is therefore proportional to $T^{(\alpha - 5/2)}.$  Thus, if $\alpha < 5/2$, as claimed by GR90 except when $m_{\rm X}/m_{\rm n} \gg 1$, the concentration of the dilute gas is a decreasing function of temperature.   This leads to the pathological result that a medium containing a binary gas mixture with a temperature gradient will undergo segregation, even in the absence of a gravitational field; moreover, the sense of segregation is the same (dilute gas concentration largest where the temperature is smallest) regardless of whether the dilute gas has a larger or smaller molecular mass than the background gas.   Such behavior is neither observed nor understandable on thermodynamic grounds. 
Finally, we note that in a gas containing multiple constituents with different molecular masses, the sum of the differential equations for the individual components is guaranteed to yield the hydrostatic equilibrium equation when equation (8) is adopted for the derivatives of the partial pressures, whereas it does not if the GR90 modification is included.  Accordingly, we adopt the density distribution implied by equation (8) in this study.

\section{Loss of HIDM from the thermosphere}

To compute the loss rate of HIDM, we may write the escaping flux as 
$$F_{\rm es}=\int_0^\infty L(z) dz, \eqno(B1)$$ where $L(z)$ is the loss rate per unit volume.  The latter may be written 
$$L(z) = \int_{v_{\rm es}}^\infty R_v\, \beta \,dv \eqno(B2)$$ where $R_v dv$ is the rate per unit volume at which scattering events produce HIDM particles with speeds of $v$ to $v+dv$, and $\beta$ is the probability that a scattered particle with velocity $v > v_{\rm es}$  actually escapes (instead of suffering an additional scattering).    For a particle traveling in the upward direction, the probability of escape is $\exp(-\tau)$, where 
$\tau = \int_z^\infty  \lambda^{-1}\, dz.$
\ma{Because the escape of HIDM is dominated by particles at or just above the escape velocity, the relevant cross-sections are those for collision velocities of $v_{\rm es}$.} 

Since $z \ll R_\earth$, we may treat the atmosphere as having a plane-parallel geometry.  For a scattered particle traveling at angle $\cos^{-1}\mu$ to the upward direction, the escape probability is reduced to $\exp(-\tau/\mu)$.  The angle-averaged escape probability is therefore $\beta\bzzz{(\tau)} = \onehalf E_2(\tau),$ where $E_2 = \int_0^1 \exp(-\tau/\mu) d\mu$ is the exponential integral function of order 2.

We may estimate $R_v$ by observing that the fraction of HIDM particles, $f_vdv$, with speeds in the range $v$ to $v+dv$ obeys the relation
$n_{\rm DM} f_v = R_v t_v,$ 
where $t_v$ is the mean time between scatterings.  Particles with $v > v_{\rm es}$ are moving faster than the typical velocities of molecules in Earth's atmosphere.  In this limit, $t_v$ may be estimated as $\lambda/v$, and thus $R_v$ may be approximated by $n_{\rm DM} f_v v / \lambda$.
With the use of this expression for $R_v$, and substituting equation (B2) into equation (B1), we obtain for the escaping flux
$$F_{\rm es}=\int_0^\infty L(z) dz =  \int_0^\infty \int_{v_{\rm es}}^\infty {n_{\rm DM} f_v v \beta \over \lambda }\, dv\,dz, \eqno(B3)$$
\bzzz{where}
$$\bzzz{{1 \over \lambda} =\sum_{\rm A} n_{\rm A}  \sigma^{\rm A}_{v}}   \eqno(B4)$$
and with the sum taken over all atmospheric nuclei.
Observing now that $dz = \lambda d\tau$, we may rewrite equation (B3) 
$$F_{\rm es}=\int_0^\infty \int_{v_{es}}^\infty \onehalf n_{\rm DM} v f_v E_2(\tau) dv d\tau. \eqno(B5)$$
Below the LSS, $f_v$ is well-approximated by the Maxwell-Boltzmann distribution function at the temperature of the atmosphere, 
$$f_{\rm MB}(T) = \biggl({m_{\rm DM} \over 2 \pi kT}\biggr)^{3/2}  4 \pi v^2 \exp(-m_{\rm DM} v^2/2kT).\eqno(B6)$$ 

We consider first the case of an isothermal atmosphere at temperature $T_0$. 
The Jeans approximation consists of assuming $f_{\rm MB}(T_0)$ above the LSS as well.  With this assumption, our treatment yields
$$F_{\rm es}=    \int_{v_{\rm es}}^\infty   v f_{\rm MB}(T_0) dv \times \onehalf \int_0^\infty  n_{\rm DM}(\tau) E_2(\tau) d\tau$$
$$ = \biggl({2kT_0 \over \pi m_{\rm DM}}\biggr)^{1/2}\biggl(1 + {m_{\rm DM} v_{\rm es}^2 \over 2kT_0}\biggr) 
\exp\biggl(-{m_{\rm DM} v_{es}^2 \over 2kT_0} \biggr) \int_0^\infty n_{\rm DM}(\tau) E_2(\tau) d\tau. \eqno(B7)$$
If $n_{\rm DM}$ varies slowly with $\tau$, then the integral over $\tau$ may be approximated by $\onehalf n(\tau=1)$, and we recover the Jean escape formula (eqn.\ 10). 

Because the gas temperature rises rapidly in Earth's thermosphere \bzzz{(i.e.\ above an altitude of $\sim 100$~km; see Figure 3)}, we have investigated whether the isothermal treatment given above is applicable.  In particular, we have considered the possibility that collisions between HIDM and hot gas particles in the thermosphere might significantly enhance the escape rate.  Because the gas density there is very small, escape from the thermosphere proceeds almost invariably by the single scattering of a HIDM particle by an atmospheric molecule or atom.  The HIDM have a characteristic energy $kT_{\rm LSS}$, while the atmospheric molecules and atoms -- which have a much larger cross-section for collisions among themselves -- still have a Maxwell-Boltzmann distribution at the local thermospheric temperature, $T(z)$.  

We consider the scattering of \bzzz{HIDM particles} with a thermospheric \be{nucleus} of mass 
$m_{\rm A}$.  If the 
scattering angle is cos$^{-1}\bzzz{\mu_{\rm sc}}$ in the center-of-mass frame, the fractional energy transfer is to the HIDM \be{if it is initially at rest} is given by equation (4) and has an angular dependence 
$f_{\rm KE} = (1 - \mu_{\rm sc}) {\bar f_{\rm KE}}$.  The HIDM particle will acquire enough energy to escape if the kinetic energy of the \be{atomic nucleus} on which it scatters exceeds a minimum value 
$$E_{\rm min} = \onehalf m_{\rm A} v_{\rm min}^2 = {\onehalf m_{\rm DM} v_{\rm es}^2 \over (1 - \mu_{\rm sc}) {\bar f_{\rm KE}}}, \eqno (B8)$$ 
where $v_{\rm min}$ is the minimum velocity corresponding to that energy.  One-half of the HIDM thereby produced are moving in the upward direction, so the rate per unit volume at 
which scattering events 
produce escaping HIDM is 
$$dR_{\rm es} = \onehalf n_{DM} n_{\rm A} \sigma_{11}^{\rm A} \int_{v_{\rm min}}^\infty v f_{\rm MB} dv d\mu_{\rm sc}, \eqno(B9)$$
where $n_{\rm A}$ is the number density of the colliding gas particles.

Using the expression given by equation (B6) for $f_{\rm MB}$, 
and integrating equation (B9) over velocity, we obtain
$$dR_{\rm es} = \pi^{-1/2} n_{\rm DM} n_{\rm A} \sigma_{11}^{\rm A} v_{\rm T} (1 + v_{\rm min}^2/v_{\rm T}^2) \, {\rm exp}\,(-v_{\rm min}^2/v_{\rm T}^2) d\mu_{\rm sc} , \eqno(B10)$$
where $v_{\rm T} = (2kT/m_A)^{1/2}$.   Finally, integrating over angle with \be{the} assumption that $\sigma_{11}^{\rm A}$ is independent of $\mu_{\rm sc}$ (i.e.\ that the scattering angles are isotropically distributed in the center-of-mass frame, \be{as appropriate for short-range interactions in the low-velocity limit}), we find
$$R_{\rm es} = 2 \pi^{-1/2} n_{DM} n_{\rm A} \sigma_{11}^{\rm A} v_{\rm T} \exp (-\onehalf m_{\rm DM} v_{\rm es}^2 / kT_{\rm eff}), \eqno(B11)$$
where $$T_{\rm eff} = 2 {\bar f}_{KE} T(z).\eqno(B12)$$

\begin{figure}
\includegraphics[width=16 cm]{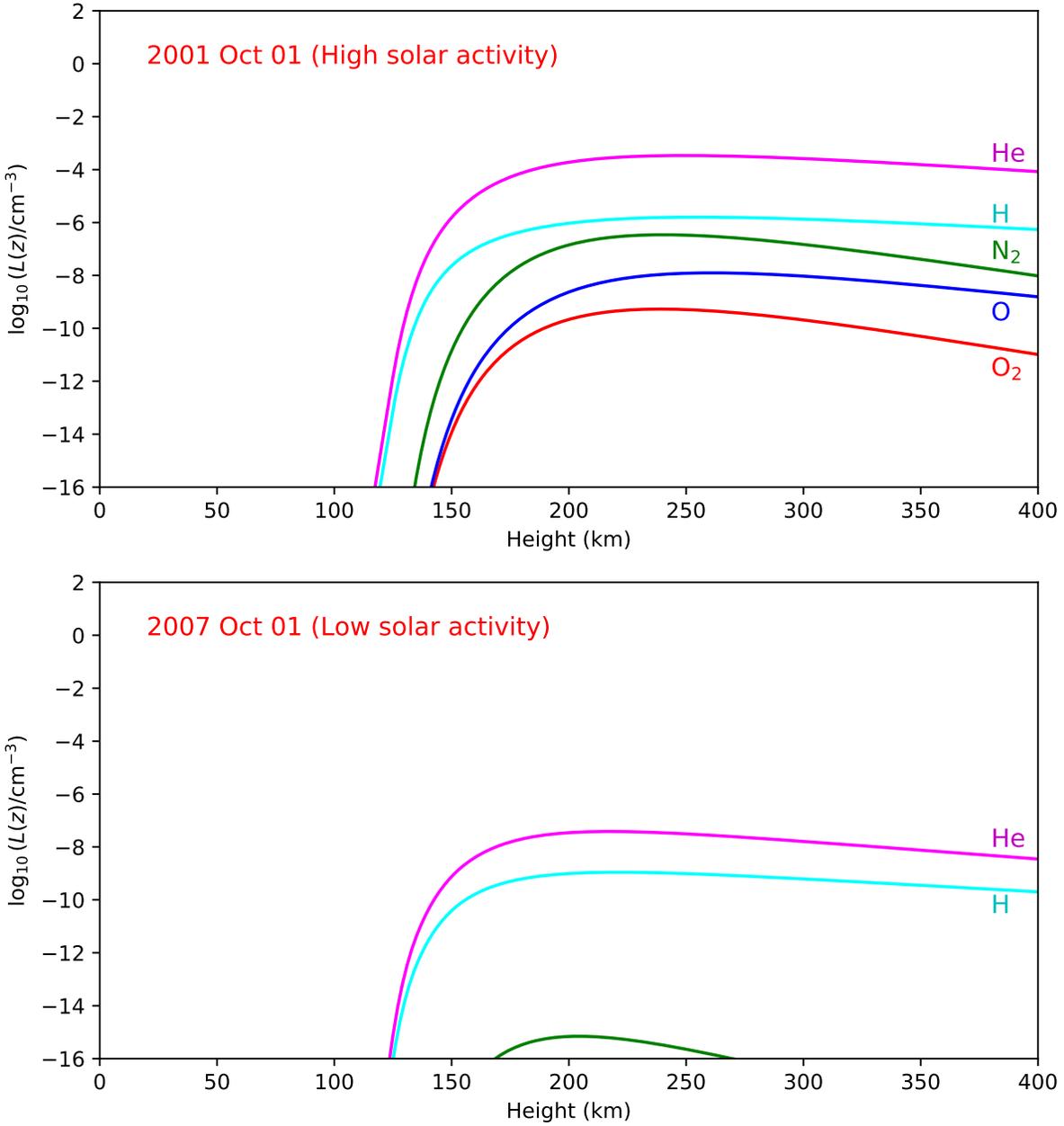}
\caption{Thermospheric loss rates per unit volume caused by collisions with different atmospheric constituents.  Results apply to m$_{\rm DM}=2\,m_{\rm p}$ and $\sigma_{11}^{\rm A}=10^{-23}\,\rm cm^2,$ and were obtained for the two atmospheric profiles plotted in Fig.\ 3}
\end{figure}

In Figure 14, we show the loss rates per unit volume, $L(z) = R_{\rm es}\beta(\tau)$, resulting from collisions with each constituent of the atmosphere.  The example results shown here were obtained for the two atmospheric profiles plotted in Figure 3 \be{in Section 2.4.1}, and for $m_{\rm DM}=2\,m_{\rm p}$ and $\sigma_{11}^{\rm A}=10^{-23}\, \, \rm cm^{2}$.  For $m_{\rm DM}=2\,m_{\rm p}$, collisions with He are the dominant loss process; despite the relatively small He abundance in the thermosphere, the energy transfer is more efficient for He (${\bar f_{KE} = 4/9}$) than for N (${\bar f_{KE} = 7/32}$), leading to a larger $T_{\rm eff}.$  
  
\subsection{Variation of the loss rate during the solar cycle}

The results shown in Figure 14 indicate that the HIDM loss rate can depend strongly on the level of solar activity.  Moreover, there are also significant dependences on latitude, local solar time, and day-of-year.  To compute the globally-averaged loss rate over an extended period, we have run a separate grid of NRLMSIS models for each day over a complete solar cycle, obtaining temperature and density profiles as a function of latitude and local solar time.  Here, we considered solar cycle 21 (1976 - 1985), for which the maximum monthly SSN (smoothed sunspot number) was the largest of any solar cycle in the $\sim 50$~years for which F10.7 indices have been available, and the second largest over the $\sim 400$~years for which sunspots have been observed.

The resultant data set provides roughly 185 million predicted temperatures - and an equal number of predicted densities for each of 8 atmospheric constituents - as a function of altitude (101 grid points from 0 to 1000 km), date (4,108 grid points), latitude (19 grid points from --90 to 90 deg), and local solar time (24 grid points from 0 to 23h UT).  For each combination of latitude, solar angle, and date, we computed the escaping flux of HIDM particles with the use of eqn.\ (B12).  
The grid points in latitude and local solar time were spaced finely enough to permit an accurate integration over the Earth's surface to obtain a globally-averaged escape flux, ${\bar F}_{\rm es},$
and the corresponding fractional loss rate, ${\bar f}_{loss}$.  Results are shown in the upper panel of Figure 15 for several values of $m_{\rm DM}$.   They exhibit large variations that result from fluctuations in solar activity.  The bottom panel of Figure 15 shows the corresponding variations in the solar activity indices.  Here, we adopted a cross-section $\sigma_{11}^{\rm A}=10^{-23}\, \rm cm^2$ for all nuclei; as indicated by equation (B11), the loss rate scales linearly with cross-section.

\begin{figure}
\includegraphics[width=13 cm]{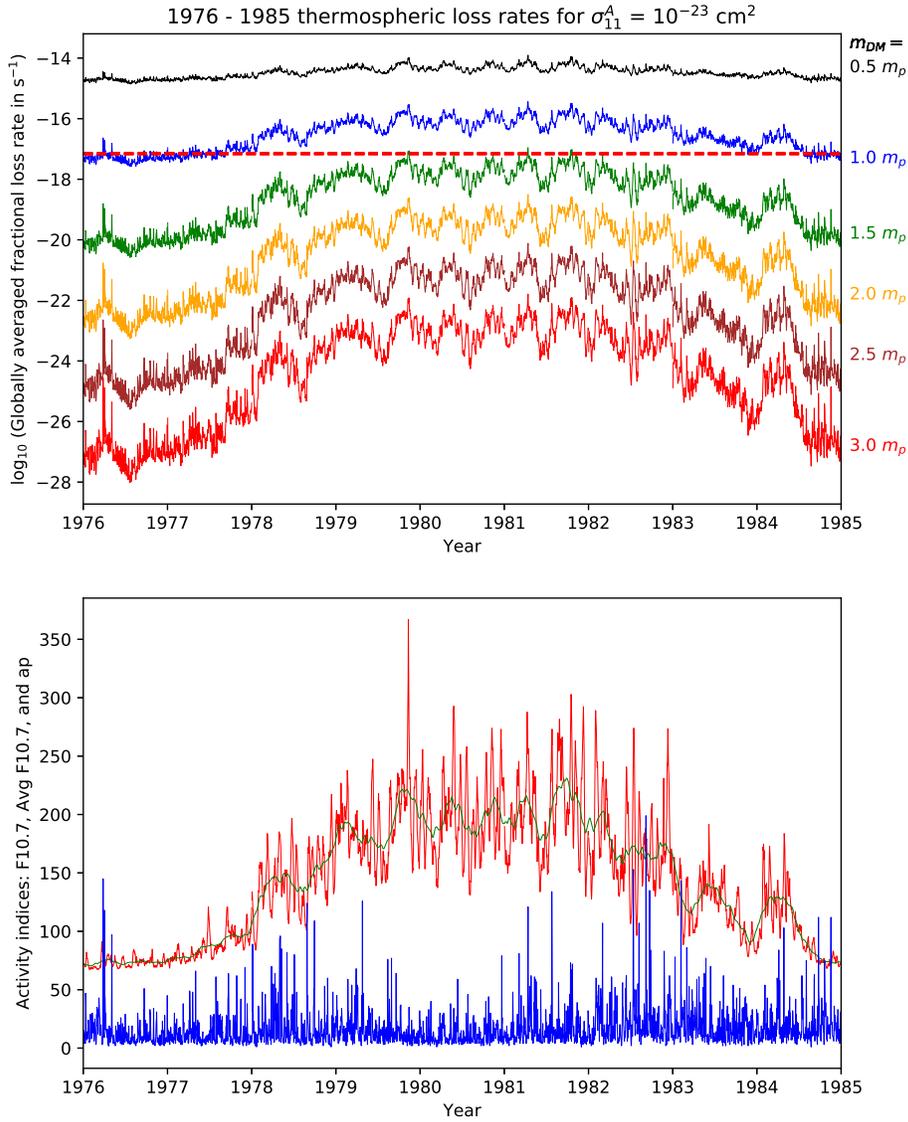}
\caption{Top panel: daily, globally-averaged, fractional loss rate over the 11-year period 1976-1985, with $\sigma_{11}^{\rm A}=10^{-23}\,\rm cm^2$ for all nuclei.
The red dashed line shows ${\bar f}_{\rm loss} = 1/t_\earth$, below which thermospheric loss is negligible.  Bottom panel: solar activity indices over the same period.  F10.7, F10.7A and $A_p$ are shown in red, green and blue, respectively.}
\end{figure}

\begin{figure}
\includegraphics[width=15 cm]{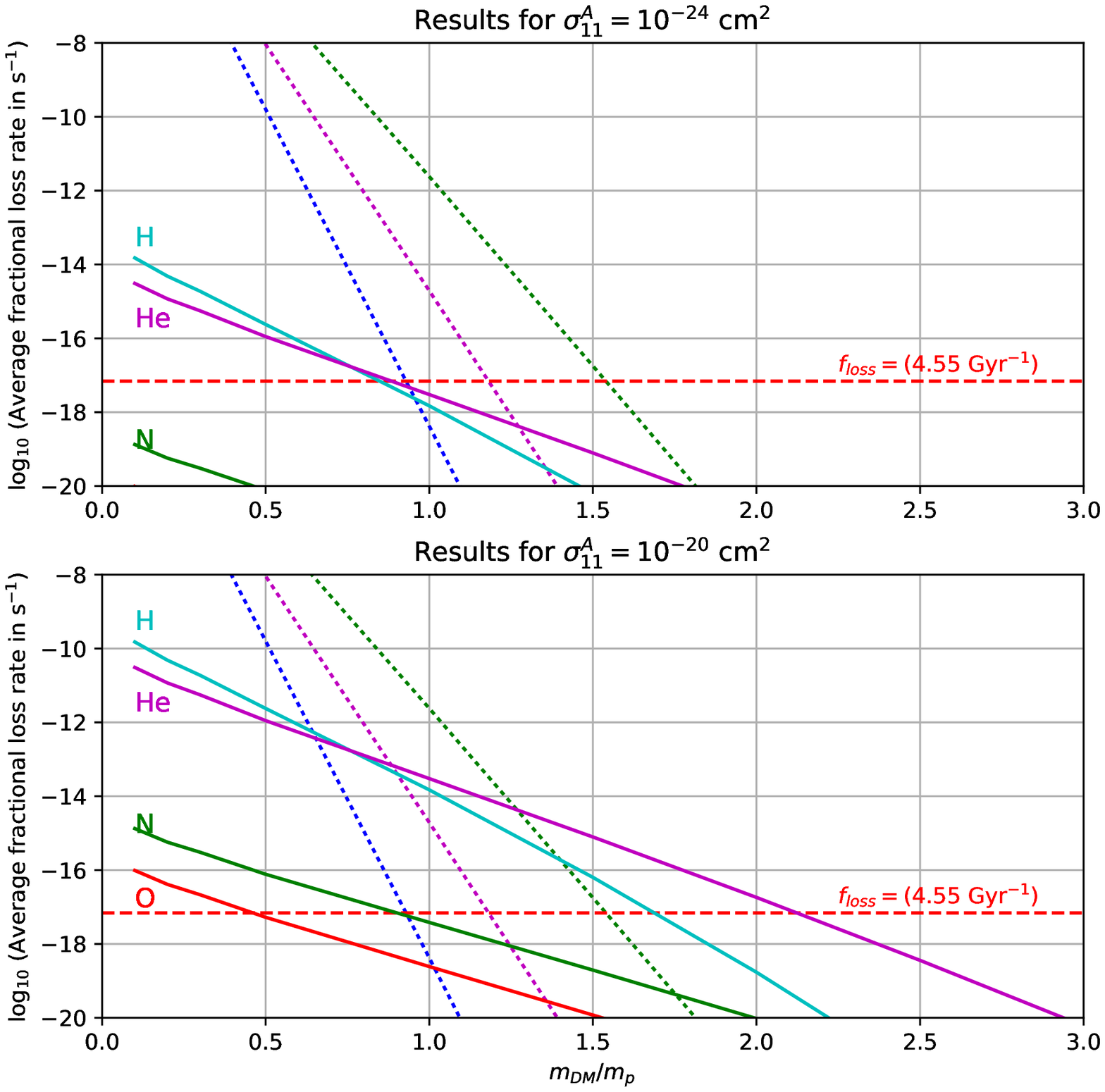}
\caption{Average thermospheric loss rates over solar cycle 21, as a function of $m_{\rm DM}$.  Green, red, magenta and cyan curves show the individual contributions due to collisions with N, O, He and H nuclei for an assumed 
$\sigma_{11}^{\rm A}$ of $10^{-24}\,\rm  cm^2$ (\be{top panel}) and $10^{-20}\,\rm  cm^2$ (\be{bottom panel}).  For comparison, the Jeans loss rates (from Figure 5) are shown by \be{dotted} curves for $\sigma_{11}^{\rm es} = 10^{-28}\,\rm cm^2$ (\be{green}), $10^{-27}\,\rm cm^2$ (\be{magenta}) and $10^{-20}\,\rm cm^2$ (blue) \be{from right to left}.}
\end{figure}

Figure 16 shows the average loss rate over the full solar cycle, as a function of $m_{\rm DM}$.  Green, red, magenta and cyan curves show the individual contributions due to collisions with N, O, He and H nuclei for an assumed 
$\sigma_{11}^{\rm A}$ of $10^{-24}\,\rm  cm^2$ (top panel) and $10^{-20}\,\rm  cm^2$ (bottom panel).
For comparison, the Jeans loss rates (from Figure 5) are shown by dotted curves for $\sigma_{11}^{\rm es} = 10^{-28}\,\rm cm^2$ (green), $10^{-27}\,\rm cm^2$ (red) and $10^{-20}\,\rm cm^2$ (blue).  The latter case (blue dotted line) represents the smallest Jeans loss rate for any value of $\sigma_{11}^{\rm es}$ considered here.  The horizontal red dashed curve indicates the value of $1/t_\earth$. \be{Figure 16 indicates that if $\sigma_{11}^{\rm H}$ and $\sigma_{11}^{\rm He}$ are both $\le 10^{-24}\,\rm  cm^2$, then thermospheric loss is never significant: the thermospheric loss rate is either smaller than $1/t_\earth$ or smaller than the Jeans loss rate.} 

\be{If either $\sigma_{11}^{\rm He}$ \bzzz{or} $\sigma_{11}^{\rm H}$ exceeds $10^{-24}\,\rm  cm^2$, then thermospheric loss can be important.  For $\sigma_{11}^{\rm He}$ and $\sigma_{11}^{\rm H}$ as large as $10^{-20}\,\rm  cm^2$, Figure 16 shows that 
the thermospheric loss rates due to collisions with H or He can exceed both the Jeans loss rate 
and $1/t_\earth$ for DM masses between $\sim 0.7$ and 2.1~$m_{\rm p}$.  
However, for $m_{\rm DM}$ in the 1.0 -- 2.1~$m_{\rm p}$ mass range, the vaporization rates for liquid H$_2$ and He (Section 3.3) place stringent limits on these cross-sections and imply that thermospheric loss is only significant over a narrower range of DM masses.  To illustrate this point, let us consider the case where $\sigma_{11}^{\rm es}=10^{-20}\,\rm cm^2$, which maximizes the importance of thermospheric loss relative to Jeans loss.  We first consider the results that are obtained in the transparent dewar limit when thermospheric loss is neglected (Figures 10 and 11). As $m_{\rm DM}$ increases from 1.0 to 2.1~$m_{\rm p}$, the limit on $\sigma_{\rm 300\,K}^{\rm He}$ increases from $0.3 - 1.3 \times 10^{-28}\,\rm cm^2$ (Figure 10), and the corresponding limit for $\sigma_{\rm 300\,K}^{\rm H}$ increases from $0.6 - 3.7 \times 10^{-28}\,\rm cm^2$ (Figure 11).  Throughout this mass range, these limits are below the cross-sections at which the dewars become opaque by a factor of at least 4000 (for He) or 1000 (for H).  This now indicates that in the {\it opaque} 
dewar limit (i.e. $\sigma_{11}^{\rm He} \simgt 6 \times 10^{-25} \, \rm cm^{2}$
or $\sigma_{11}^{\rm H} \simgt 4 \times 10^{-25} \, \rm cm^{2}$), the values of $n_{\rm DM}$ are constrained to lie at least 3 orders of magnitude below the values that are obtained without the inclusion of thermospheric escape.  Referring now to Figure 16, we see that the thermospheric loss rates for H and He never exceed $1/t_\earth$ by a factor as large as 1000 for $m_{\rm DM} \ge 1.2\,m_{\rm p}$.  Thus, for the mass range $m_{\rm DM} = 1.2 -2.1\, m_{\rm p}$, the effects of thermospheric escape fail to reduce $n_{\rm DM}$ by a factor that is sufficient to evade the constraints implied by the vaporization rates for liquid H$_2$ and He, provided that the cross-sections for H and He at $v_{\rm es}$ are no larger than those at 300~K and are both $\le 10^{-20}\,\rm cm^2$ .  Our conclusion, then, is that thermospheric loss is potentially important only within a fairly narrow range of HIDM masses: $0.7 \, m_{\rm p} \simlt m_{\rm DM} \simlt 1.2 \, m_{\rm p}.$}

\section{Measurements of temperature gradients in boreholes}

Temperature gradients have been measured within the crust in more than 30,000 boreholes \bu{widely distributed} over the surface of the Earth.  Combined with laboratory measurements of the thermal conductivity of crustal rocks, these temperature gradient measurements have been used to obtain \bzzz{an} estimate of the total power transported upwards through the crust: $P_\earth= (47 \pm 2)$~TW (Davies \& Davies 2010).

\begin{figure}
\includegraphics[width=13 cm]{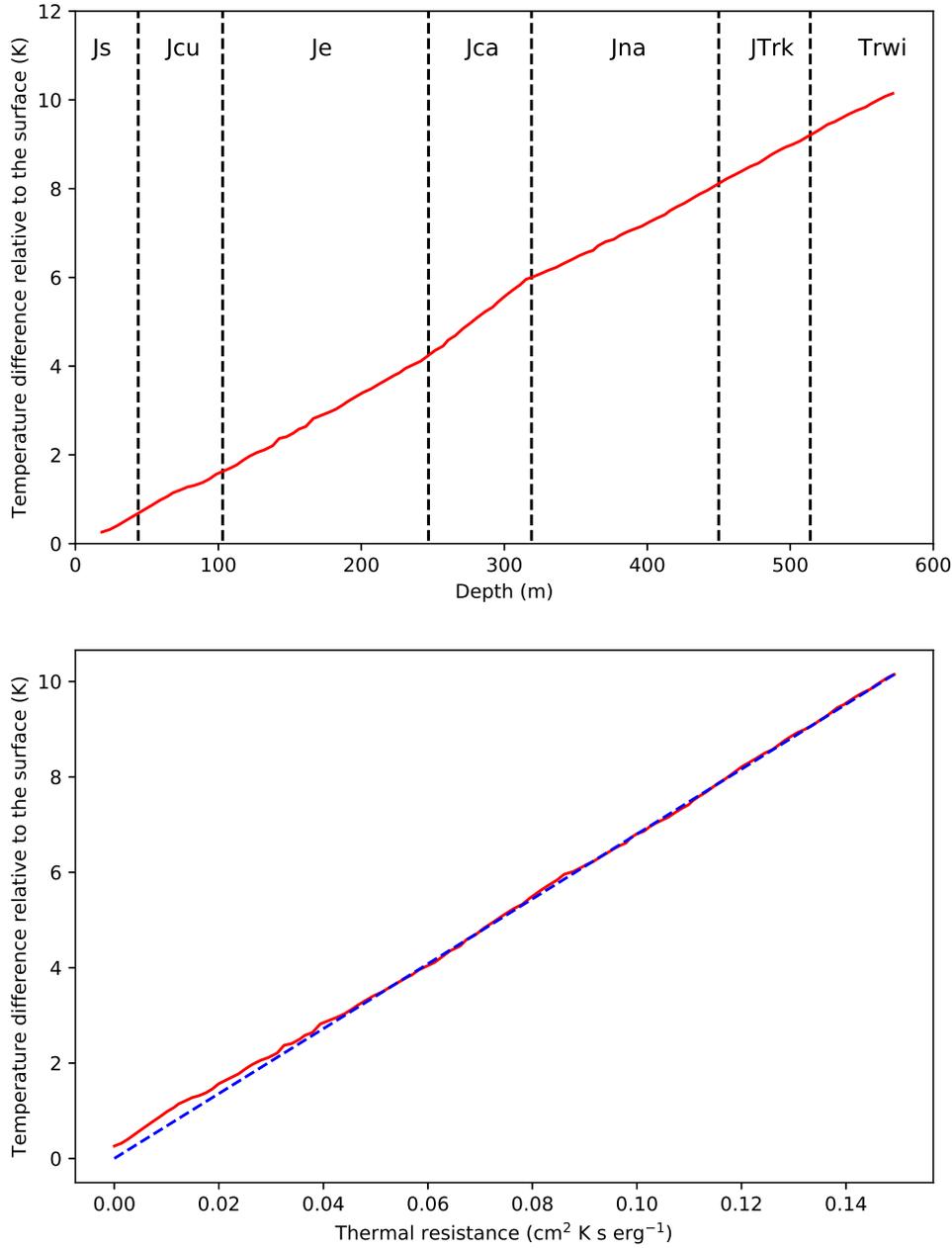}
\caption{Top panel: temperature profile measured by HC95 for the borehole designated WSR-1, adapted from their Figure 2.  Dotted vertical lines indicate the boundaries between different rock formations, and the rock types for the various layers are indicated using the abbreviations listed in Table 1 in HC95.  Lower panel: Bullard plot for WSR-1 (see the text). The dotted blue line is the behavior expected for a constant flux of $68 \rm \, erg \,s^{-1} \, cm^{-2}$. }
\end{figure}

In certain cases, boreholes cross stratified rock formations in which rock type changes.  For such boreholes, discontinuities in the measured temperature gradient can be detected at the rock formation boundaries and have been attributed to differences in the conductivity of the rock types.   As an example, we considered the study of such effects presented by HC95 for 9 boreholes in the Colorado plateau of Eastern Utah.  Changes in the temperature gradient at rock type boundaries are most obvious for the borehole designated WSR-1, in which a layer of lower- conductivity rock (Jurassic Carmel, Jca) lies sandwiched between two rock types of high conductivity (Jurassic Entrada, Je; and Jurassic Navajo, Jna).  The temperature profile for this borehole is shown in Figure 17 (top panel).   Using a simultaneous fit to all the data acquired for the nine boreholes that they investigated, HC95 obtained estimates of the different rock conductivities needed to account for the observed changes in temperature gradient at the rock formation boundaries, and compared them with laboratory-measured values.  Because the actual heat flux is not measured directly, only the relative conductivities are constrained.  For the best-fit thermal conductivities obtained by HC95, the bottom panel of Figure 17 shows a ``Bullard plot" (Bullard 1939), in which the temperature is plotted as a function of the thermal resistance,
$$ R = \int_0^y k_{\rm rock}^{-1} dy^\prime, \eqno(C1) $$ 
where $y$ is the distance below the surface.  The absence of slope discontinuities at the layer  boundaries in the lower panel of Figure 17 indicates that the relative conductivities have been computed correctly.  Moreover, the absence of significant curvature implies that the flux is constant and that there is little radiogenic heat production.  The dotted blue line is the behavior expected for a constant flux of $68 \rm \, erg \,s^{-1} \, cm^{-2}$.   The deviation from this behavior at smaller depths ($\le 100$~m) was interpreted by HC95 as providing a record of surface temperature changes over the past several hundred years.

In Figure 18, we plot the laboratory-measured rock conductivities, $k_{\rm lab}$, as a function of the values inferred from the temperature profile within the sample of boreholes, $k_{\rm bh}$.  These values are taken from Table 1 in HC95, where they are denoted $k_{\rm pr}$ and $k_{\rm a}$ respectively, and they apply to the San Rafael Swell region, in which four of the nine boreholes are located (including WSR-1).  
The error bars on $k_{\rm lab}$ represent the standard deviations obtained for laboratory measurements performed on multiple samples of a given rock type.    \bu{While the ratios of the $k_{\rm bh}$-values reported by HC95 are determined by the borehole measurements, the overall scaling adopted by HC95 is unconstrained.  (It was chosen by HC95 to minimize the differences between $k_{\rm lab}$ and $k_{\rm bh}$ when averaged over all rock types.)  Accordingly, we performed the linear regression shown by the \be{black} line with both the slope and y-intercept unconstrained, and indeed recovered a best-fit slope of 1.0}

\begin{figure}
\includegraphics[width=15 cm]{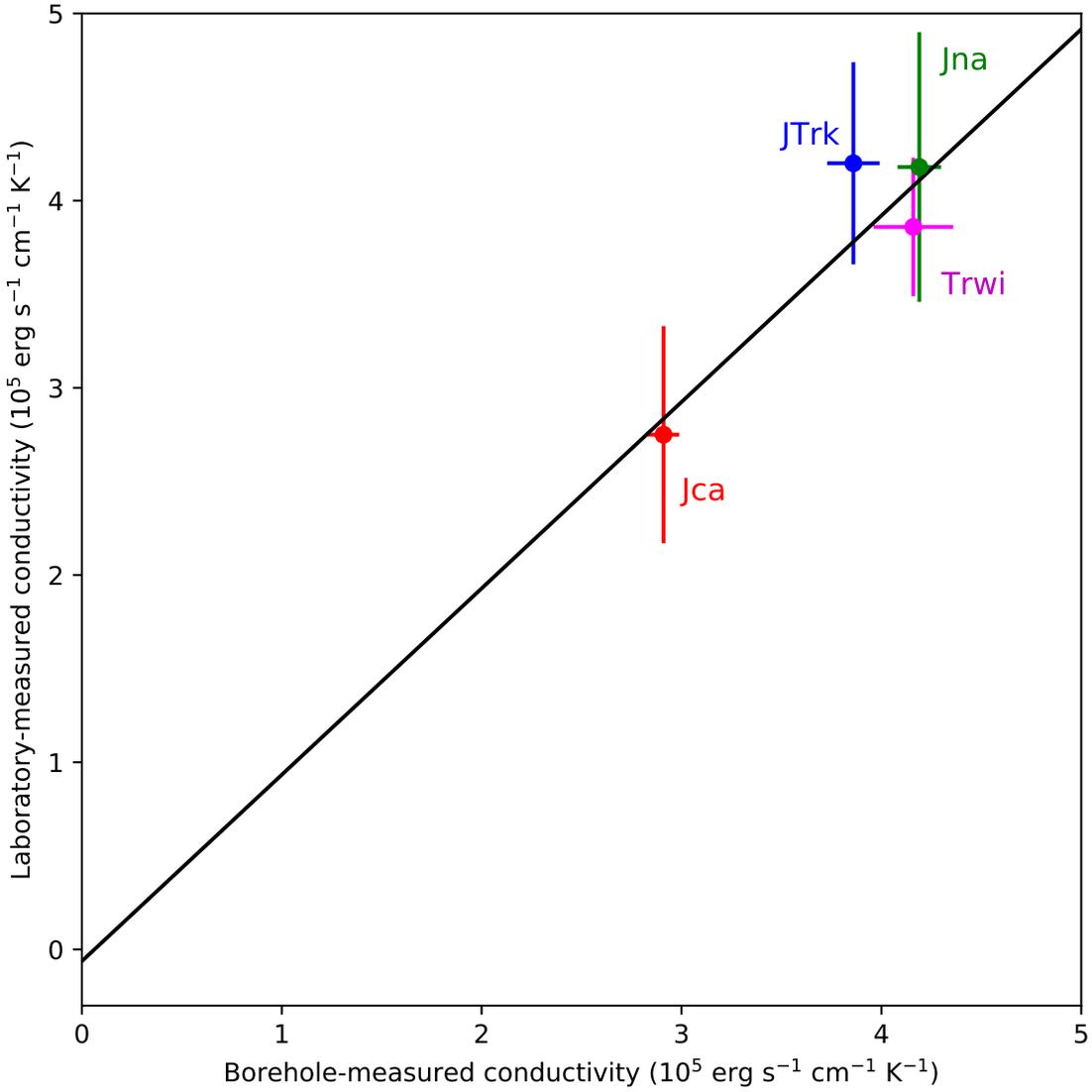}
\caption{Laboratory-measured conductivities, $k_{\rm lab}$, as a function of those measured in boreholes, $k_{\rm bh}$, for four rock types studied by HC95.  Here, we use the abbreviations listed in Table 1 in HC95 for the different rock \bu{types}.}
\end{figure}

For conductivities in the range of interest, the mean-free path for HIDM, $\lambda$, is larger than the thickness of the rock samples for which the laboratory measurments were obtained ($\sim 2\, \rm cm$; Roy et al.\ 1968).  Any anomalous conductivity, $k_{\rm DM}$, associated with HIDM would therefore increase $k_{\rm bh}$ but not $k_{\rm lab}$.  Thus, in this regime, $k_{\rm lab} = (k_{\rm bh} - k_{\rm DM}$), and thermal conductivity associated with HDIM would reveal itself as a negative intercept for the best-fit linear regression.  The linear regression yields a $y$-intercept of $(-0.04 \pm 1.44) \times 10^5\,\rm erg\,s^{-1}\,cm^{-1}\,K^{-1},$ placing a $3\,\sigma$ upper limit of  
$4.3 \times 10^5\,\rm erg \,s^{-1}\,cm^{-1}\,K^{-1}$ on $k_{\rm DM}$.  \bu{The corresponding upper limit on $\lambda$ is} $250 (m_{\rm DM}/m_{\rm p})^{1/2} \bu{(T_{\rm DM}/{\rm 300\, K})^{-1/2}} (n_{\rm DM} / 10^{14}\,\rm cm^{-3})^{-1} \rm cm$, \bu{and the resultant lower limit on $\sigma^{\prime \rm cr}_{\rm 300 K}$ is $5.4 \times 10^{-26} (m_{\rm DM}/m_{\rm p})^{-1/2} \bu{(T_{\rm DM}/{\rm 300\, K})^{1/2}} (n_{\rm DM} / 10^{14}\,\rm cm^{-3}) \rm \,cm^2$}.

\end{document}